\pdfoutput=1
\documentclass[%
reprint,
superscriptaddress,
 amsmath,amssymb,
 aps,
 showkeys,
 longbibliography,
]{revtex4-2}

\usepackage{caption}
\usepackage{subcaption}
\usepackage{tikz}
\usepackage{pgfplots}
\usetikzlibrary{shapes}
\usetikzlibrary{plotmarks}
\usepackage{amsmath}
\usepackage{graphicx}
\usepackage{dcolumn}
\usepackage{bm}
\usetikzlibrary{calc}

\pgfplotsset{compat=1.18}

\begin{document}
\title{Assessment of a Multiphase Formulation of One-Dimensional Turbulence using Direct Numerical Simulation of a Decaying Turbulent Interfacial Flow}

\author{A.~Movaghar}
 \affiliation{Department of Applied Mechanics and Maritime Sciences, Chalmers University of Technology, Gothenburg, Sweden
}%
\author{R. Chiodi}%
\affiliation{ %
Sibley School of Mechanical and Aerospace Engineering, Cornell University, Ithaca,
New York 14853, USA
}%

\author{M. Oevermann}
\email[Author to whom correspondence should be addressed: ]{michael.oevermann@b-tu.de}

\affiliation{
  Institut f\"ur Mathematik,
  Brandenburgische Technische Universit\"at Cottbus-Senftenberg, Cottbus, Germany}%
\affiliation{Department of Applied Mechanics and Maritime Sciences, Chalmers
University of Technology, Gothenburg, Sweden}

\author{O. Desjardins}
\affiliation{ %
Sibley School of Mechanical and Aerospace Engineering, Cornell University, Ithaca,
New York 14853, USA
}%

\author{A. R. Kerstein}
\affiliation{%
Consultant, 72 Lomitas Road, Danville, CA 94526, USA
}%

\begin{abstract}
The interaction between turbulence and surface tension is studied numerically using the one-dimensional-turbulence 
(ODT) model. ODT is a stochastic model simulating turbulent flow evolution along a notional one-dimensional
line of sight by applying instantaneous maps that represent the effects of individual turbulent eddies on property
fields. It provides affordable high resolution of interface creation and property gradients within each phase, which 
are key for capturing the local behavior as well as overall trends, and has been shown to reproduce the main features 
of an experimentally determined regime diagram for primary jet breakup. Here, ODT is used to investigate the interaction of turbulence with an initially planar interface. The notional flat interface is inserted into a periodic box of decaying homogeneous isotropic turbulence, simulated for a variety of turbulent Reynolds and Weber numbers. Unity density and viscosity ratios serve to focus solely on the interaction between fluid inertia and the surface-tension force. Statistical measures of interface surface density and spatial structure along the direction normal to the initial surface are compared to corresponding direct-numerical-simulation (DNS) data. Allowing the origin of the lateral coordinate system to follow the location of the median interface element improves the agreement between ODT and DNS, reflecting the absence of lateral non-vortical displacements in ODT. Beyond the DNS-accessible regime, ODT is shown to obey the predicted parameter dependencies of the Kolmogorov critical scale in both the inertial and dissipative turbulent-cascade sub-ranges. Notably, the probability density function of local fluctuations of the critical scale is found to collapse to a universal curve across both sub-ranges.

\end{abstract}

\maketitle

\begin{quotation}

\end{quotation}

\section{Introduction}
The interface/turbulence interaction between two fluids in a turbulent environment has an important role in many technical processes, {{e.g.~}}spray painting {{and}} primary liquid atomization in combustion 
devices. Primary atomization has {{a}} significant role in spray formation and its characteristics. Combustion performance such as {{efficiency}} and emissions creation is extremely dependent on spray characteristics.
{{For these reasons}}, primary atomization has been studied theoretically and experimentally for a long time \cite{linne2013imaging, desjardins2013direct, herrmann2010detailed}{{, but}} our understanding {{of}} liquid atomization is still inadequate.

Atomization in turbulent environments involves a vast range of length and time scales. Predictive simulations with high spatial and temporal resolution, i.e. direct numerical simulations (DNS) and high resolution large-eddy simulations
(LES), are used to study liquid-gas interface dynamics during primary breakup, but resolution of all relevant scales is limited by the available computational resources \cite{menard2007coupling, gorokhovski2008modeling, lebas2009numerical}. 

Consequently, for practical simulations of engineering interest as well as to investigate
the physics and scalings of primary breakup beyond the parameter range of DNS
studies, a predictive and computationally affordable interface dynamics model is highly desirable. 

Past studies showed that the surface instability on the liquid jet core has a critical role in the jet breakup process. These instabilities are  interpreted mainly by using linear analysis \cite{sallam2002liquid, ben2006assisted, marmottant2004spray, dumouchel2008experimental}, but recent studies show {that the assumptions of linear stability analysis are violated in the presence of a significant non-zero normal velocity at the interface. Therefore such a formulation is not sufficient for describing atomizing liquids in complex geometries or at high Reynolds numbers. 

So other approaches are needed to understand and model the interactions between two immiscible fluids in a turbulent environment. Such turbulence-interface interactions have been studied by several researchers.

Li and Jaberi \cite{li2009turbulence} studied the interplay between surface tension and baroclinity near the interface and their impact on turbulent kinetic energy dissipation. Trontin {\it et al.} \cite{trontin2010direct} isolated the interaction between fluid inertia and surface tension in a box of{three-dimensional decaying homogeneous turbulence and studied anisotropic effects of
surface tension on the surrounding turbulence. Studies conducted by McCaslin {\it et al.} \cite{mccaslin2014theoretical, mccaslin2015development} for a case similar to \cite{trontin2010direct} showed that surface tension increases energy in the flow field at small scales and that interface corrugations are greatly suppressed at length scales smaller than the critical radius at which there is a balance between strain-induced wrinkling and surface-tension-induced smoothing of corrugations.

One objective of this study is to support the development of a new modeling approach for turbulent jet breakup based on the one-dimensional-turbulence (ODT) model. ODT was recently used \cite{movaghar2017numerical} to reproduce the main features of an experimentally determined regime diagram for primary jet breakup, followed by detailed comparison to a three-dimensional numerical simulation \cite{movLinHerKer2018}. This stochastic modeling approach provides high resolution affordably by resolving all relevant scales only in the direction normal to the phase interface using a modeling construct that captures three-dimensional effects. The low computational cost of ODT compared to fully resolved three-dimensional DNS overcomes the restriction of DNS to moderate Reynolds and Weber numbers.

Here, the ODT model representation of phase-interface motion within turbulent flow is assessed by comparing ODT and DNS results for a regime that isolates just the interactions between surface tension and turbulence. In both DNS and ODT simulations, a planar interface is inserted into a box of decaying homogeneous isotropic turbulence (HIT) and the flow-induced deformation of the initially planar interface and response of the flow to the associated surface-tension effects is examined.

The DNS time advances the incompressible Navier-Stokes equations for immiscible two-phase flow in which all fluid properties are considered to be constant 
and identical in the two phases. For this case, the momentum equation is given by
\begin{equation}
  \dfrac{\partial {\pmb{u}}}{\partial t} + {\pmb{u}} \cdot \nabla {\pmb{u}} = \dfrac{1}{\rho}\nabla p  + \nu \nabla^{2} {\pmb{u} + \dfrac{1}{\rho} \pmb{f}_{\sigma} }, 
\end{equation}
where ${\pmb{u}}$ is the velocity, $\rho$ the density, $p$ the pressure, $\nu$ the kinematic
viscosity and $\pmb{f}_{\sigma}$ the surface tension force, which is nonzero only at the phase interface.

\section{Flow configuration}

The studied flow configuration was previously investigated using DNS \cite{mccaslin2014theoretical,mccaslin2015development}, albeit for different values of the governing parameters than those used in the present study, which are shown in Table \ref{tab:table2}. It involves two stages.

First, the free decay of turbulence is simulated until it reaches the homogeneous isotropic state. After reaching HIT, an interface is inserted at the targeted value of the Taylor-microscale Reynolds number $ Re_{\lambda_{g}} = u_{rms}\lambda_{g}/\nu$, where $u_{rms}$ denotes the root-mean-square velocity fluctuation and $\lambda_{g} = \sqrt{10}(\eta^{2}L_{int})^{1/3}$ is the Taylor microscale. Here, $\eta$ is the Kolmogorov length scale and $L_{int}$ is the characteristic length scale of the large eddies. In terms of the turbulent kinetic energy (TKE) $k$ per unit mass, which is $k = \frac{3}{2}u_{rms}^{2}$ for isotropic turbulence, and the TKE dissipation rate $\epsilon$, these length scales are $\eta = (\nu^3/\epsilon)^{1/4}$ and $L_{{{int}}} = k^{3/2}/\epsilon$. In order to focus on turbulence-interface interactions, the same density and viscosity are assigned for both phases and a phase index is used to distinguish them.

When turbulence has decayed to a prescribed Taylor-scale Reynolds number $Re_{\lambda_{g}}$, which in this study is $155$ (except for one ODT case for $Re_{\lambda_{g}} = 500$), the second part of the simulation is initiated by inserting a planar phase interface into the middle of the box. The phase on each side of the interface is assigned a unique index. In data reduction involving time-varying flow parameters, their values at the instant of interface insertion at nominal time $t=0$ are used because the physical meaning of these becomes ambiguous as surface-tension effects induce flow inhomogeneity.

As indicated in Table \ref{tab:table2}, ODT results are reported for a range of values of surface tension $\sigma$, corresponding to a range of turbulent Weber numbers, defined as $W\!e_{\lambda_{g}} = \rho u_{rms}^{2} \lambda_{g}/\sigma $, evaluated at the instant of interface insertion. Model results are compared to the available DNS cases.

\begin{table}
\caption{Cases simulated using ODT. Cases 1-4 were also simulated using DNS.}
\begin{ruledtabular}
\begin{tabular}{cccc}
 Case&$Re_{\lambda_{g}}$&$W\!e_{\lambda_{g}}$&\\
\hline
1&  155 & $ \infty$\\
2&  155 & 1.36  \\
3&  155 & 8.47  \\
4&  155 & 21.06 \\
5&  155 & 100 \\
6&  155 & 500 \\
7&  155 & 1000  \\
8&  155 & 4000  \\
9&  155 & 8000 \\
10& 500 & 100 \\
\end{tabular}
\end{ruledtabular}
\label{tab:table2}
\end{table}

\section{Methods}

\subsection{One-dimensional turbulence}

The ODT formulation used in this study is described briefly in Appendix \ref{Sec:ODT}. For detailed explanations, see the overview by Kerstein \cite{kerstein2022} and extensions of the approach by Ashurst and Kerstein \cite{ashurst2005one} and by Movaghar {\it et al.} \cite{movaghar2017numerical, movaghar2018modeling}. The advantages of a turbulence model formulated as a one-dimensional unsteady stochastic simulation are twofold. First, a one-dimensional formulation enables affordable simulation of high-Reynolds-number turbulence over the full range of dynamically relevant length scales, {{resolving}} the {{interactions between}} turbulent advection and microphysical processes such as viscous dissipation. Second, this approach permits high resolution of properties in the direction of the most significant gradients or flow-structure variations, here denoted by the spatial coordinate $y$, and on that basis is found to reproduce diverse flow behaviors.

In contrast to common approaches based on the Navier-Stokes equations, ODT uses a set of mechanisms modeling the physical effects phenomenologically on a 1D line of sight through the domain. The property fields defined on the one-dimensional domain evolve by two mechanisms: molecular evolution and a stochastic process representing advection. The stochastic process consists of a sequence of `eddy events,' each of which involves an instantaneous transformation of the velocity and any other property fields. In this application the only other property field is the phase index, which is subject to advection but not molecular evolution. During the time interval between each event and its successor, molecular evolution occurs, governed by the equation
\begin{align}
{\partial u_{i}(y,t)}/{\partial t} = {{\nu}}{\partial^{2} u_{i}(y,t)}/{\partial y}^{2},
\label{momentum1}
\end{align} 
where $u_{i}$ with $i \in {1,2,3}$ are the three velocity {{{components}}}.

The eddy events representing advection may be interpreted as the model analog of individual turbulent eddies. In ODT each eddy event is characterized by a length scale (the eddy size) and a time scale. The stochastic sampling of occurrences of eddies is based on the interpretation of the eddy time scale as the mean time until the next occurrence of an eddy of given size at a given location. The unique feature is that the time scale is based on the instantaneous flow state within the spatial extent of the eddy, so it is different for each sampled eddy rather than being based on a mean-field relationship. Thus, the velocity profiles $u_{i}(y,t)$ do not advect fluid along the $y$ coordinate, but they indirectly influence advection through their role in determining the time scale of individual eddies, and thus the time-varying rates of occurrences of various eddies. This enables the model to capture dynamical details that are flow specific based on the initial and boundary conditions and any local or distributed energy sources and sinks, such as surface tension, whose interaction with turbulent flow is the present focus.

The instantaneous time scale governing the sampling of a given eddy is obtained using the appropriate dimensional combination of the eddy size and a measure of kinetic energy based on the profiles $u_{i}(y,t)$ within the eddy spatial interval. This approach allows surface tension (and other such effects) to be incorporated by modifying the kinetic energy due to eddy-induced change of the total phase interface associated with the eddy. This modification is implemented by changing the profiles $u_{i}(y,t)$ within the eddy interval in a way that applies the prescribed kinetic-energy change while conserving the $y$-integrated momentum of all velocity components.

This procedure has two effects. First, it modifies the likelihood of eddy occurrence during a given time increment. For example, this likelihood is zero if the surface-tension effect requires a kinetic-energy reduction that exceeds the presently available kinetic energy within the eddy interval, indicating that the eddy is energetically forbidden. Second, if the eddy is implemented, the associated changes of the profiles $u_{i}(y,t)$ represent the surface-tension-induced flow modification. Thus, eddy events not only advect the phase index, resulting in an implied change of interface area, but they also capture the effect of the latter on the flow.

Periodic boundary conditions are applied in all directions (therefore for ODT, in the $y$ direction, which is the only available direction). When fluid crosses a periodic boundary in the $y$ direction, its phase index flips so that the periodicity does not cause the creation of artificial phase interfaces.

As shown in Appendix \ref{Sec:ODT}, eddy events can increase but not decrease interface surface area and ODT as presently formulated has no mechanism representing the latter. For the decaying HIT configuration considered here, this implies that the eventual restoration of interface planarity after the turbulent motions are fully dissipated cannot be captured by the model. Therefore the present study focuses on the early development of surface area, which is the regime of strongest interaction between turbulence and surface tension and hence is of particular interest. Accordingly, the time duration of simulated realizations is short enough so that the physical interface never reaches the lateral boundaries during DNS runs.

Reported results for all ODT cases are ensemble averages of 2000 simulated realizations. Due to run-to-run statistical variability, the physical interface reaches a $y$ boundary during a few of these realizations. Then the phase indices are changed as needed to prevent the interface from crossing the boundary and thus appearing unphysically near the opposite boundary due to the periodic boundary conditions. Owing to the rarity of this situation, the impact of this modification on the statistical outputs that are reported is negligible.

\subsection{Direct numerical simulation}

The DNS data is generated using a full three-dimensional incompressible Navier-Stokes flow solver~\cite{DesjardinsNGA2008,Desjardins2013}. Each phase is transported using an unsplit geometric semi-Lagrangian volume of fluid (VOF) method~\cite{Owkes2014}, with the curvature calculated through a mesh decoupled height function~\cite{Owkes2015} and the pressure jump due to surface tension is imposed using the ghost fluid method~\cite{Fedkiw1999}.

\section{Validation of the ODT representation of HIT} \label{Valid}

To simulate HIT, the ODT flow state is initialized with a low-wavenumber narrowband velocity profile. Periodic boundary conditions are imposed. $u_{rms}$ is initially high enough so that $Re_{\lambda_{g}}$ is much greater than the target value. As a result, ODT relaxes to a state that corresponds to freely decaying HIT when this target value is reached. A planar phase interface is then inserted at the midpoint of the domain, followed by further time advancement. The instant of interface insertion is designated as time $t=0$, and the time coordinate is scaled in all plotted results by the large-eddy turnover time $\tau = k / \epsilon$ evaluated at $t=0$. 

ODT model parameters were set by comparison to the infinite-$W\!e_{\lambda_{g}}$ DNS case, for which the flow continues to behave as decaying HIT after $t=0$ because the interface is dynamically passive for this case. Parameter setting is based only on the flow properties for this case, so interface evolution for this case, as well as for finite-$W\!e_{\lambda_{g}}$ cases, is a model prediction that is shown in Section \ref{sec:interface}. On this basis, the ODT model parameters defined in Appendix \ref{sec:selec} are chosen to be $ C = 5.2$ and $ Z = 10$.

$C$ scales the overall eddy rate, and therefore the turbulence decay rate. Figure \ref{TKEhistory} shows the evolution of turbulent kinetic energy in both ODT and DNS simulations of decaying HIT. The good agreement of ODT with the DNS turbulence decay was obtained by tuning $C$ to match the DNS energy dissipation at $t=0$.

\begin{figure}[!ht]
\begin{center}
\begin{tikzpicture}
 \node[inner sep=0pt] (A) {\includegraphics[width=2.5in]{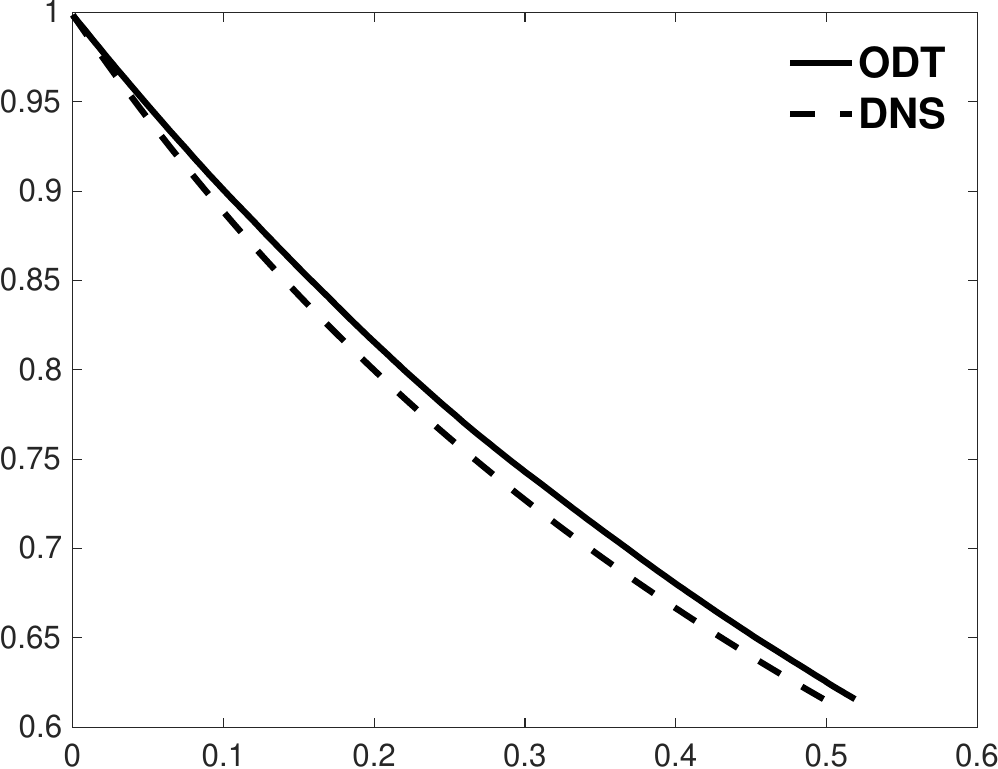}};
  \node[black] (B) at ($(A.south)!-.09!(A.north)$) {$t / \tau$};
  \node[black,rotate=90] (C) at ($(A.west)!-.03!(A.east)$) {$TKE(t)/TKE(0)\ \ $};
\end{tikzpicture}
\end{center}
\caption{Turbulent kinetic energy evolution for homogeneous decaying turbulence}
\label{TKEhistory} 
\end{figure}

$Z$ scales the viscous suppression of eddy occurrences, mainly at small scales, and therefore primarily affects the Kolmogorov scale. Importantly, the ODT treatment of viscous processes captures key features of the viscous-inertial balance such as the correct $Re_{\lambda_{g}}$ dependence of the wavenumber ($k$) range of the inertial cascade.

The influence of $Z$ is thus seen in the extent of the inertial range for given $Re_{\lambda_{g}}$ prior to the high-wavenumber roll-off of the velocity spectrum into the dissipative wavenumber range. The comparison of ODT and DNS spectra is shown in Fig.~\ref{spectrum}.

\begin{figure}[!ht]
\begin{center}
\begin{tikzpicture}
  \node[inner sep=0pt] (A) {\includegraphics[width=2.5in]{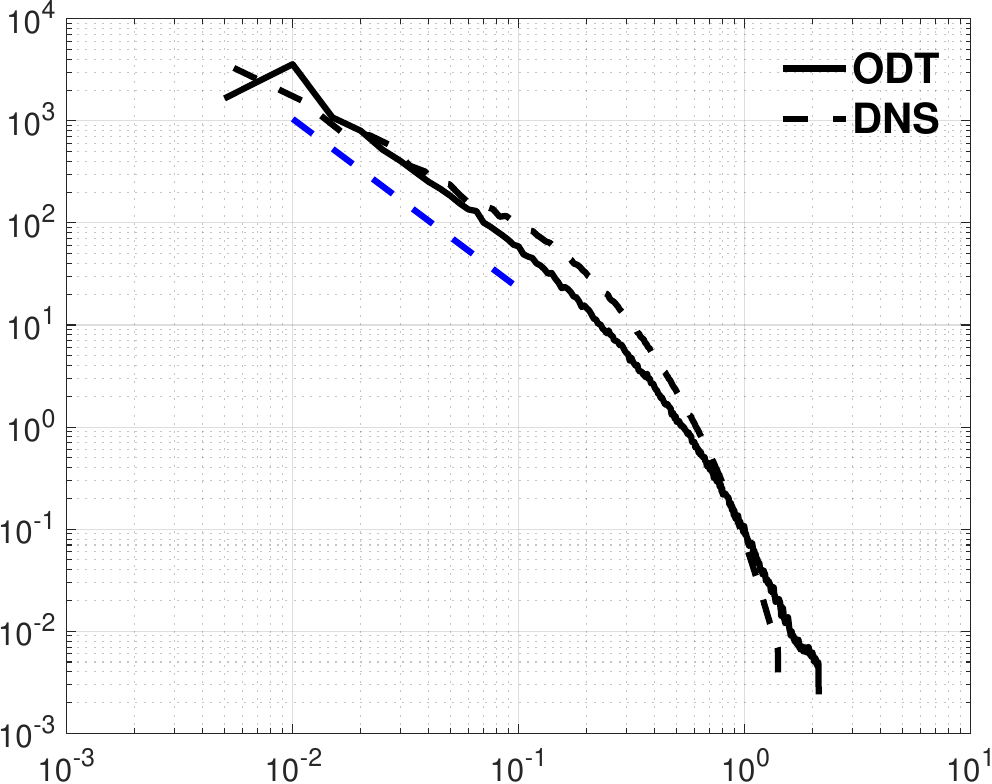}};
  \node[black] (B) at ($(A.south)!-.09!(A.north)$) {$k\eta$};
  \node[black,rotate=90] (C) at ($(A.west)!-.03!(A.east)$) {$E_{{22}}(k)/(\epsilon\nu^{5})^{1/4}$};
\end{tikzpicture}
\end{center}
\caption{For homogeneous decaying turbulence at $Re_{\lambda_g} = 155$, normalized DNS and ODT one-dimensional spectra of transverse velocity fluctuations $E_{{22}}(k)/(\epsilon\nu^{5})^{1/4}$ (black lines); $(k\eta)^{-5/3}$ (dashed blue line).} 
\label{spectrum} 
\end{figure}

In ODT, the distinction between the spectral properties of longitudinal and transverse velocity components is not captured because these velocities do not directly advect fluid in ODT and are not subject to the solenoidal condition that creates the distinction between longitudinal and transverse velocity in physical turbulence.  (Alhough the solenoidal condition has no meaning in ODT, the 1D analogs of this and other conservation properties are obeyed by ODT.) Therefore the ODT one-dimensional velocity spectra $E_{jj}(k)$ are the model analogs of the transverse velocity spectrum $E_{22}(k)$ for all $j$. This spectrum, normalized by $ (\epsilon\nu^{5})^{1/4}$, is shown in figure \ref{spectrum}. A $k$ interval exhibiting the $k^{-5/3}$ inertial-range scaling is seen. As explained previously \cite{kerstein1999one}, this scaling is an outcome reflecting ODT conservation properties and multiplicative (and therefore self-similar) scale reduction by triplet maps, rather than a behavior that is hard-wired into the model.

Because ODT and DNS are compared in Fig.\ \ref{spectrum} at the same value of $Re_{\lambda_{g}}$ and $Z$ has been tuned to match the $k$ range of the DNS inertial cascade, the degree of consistency that is seen in the figure is expected. A noteworthy feature is that the ODT inertial range is in better conformance to $k^{-5/3}$ scaling than is the DNS spectrum. It is known that higher $Re_{\lambda_{g}}$ is needed to see close conformance of DNS spectra with this scaling. (The chosen $Re_{\lambda_{g}}$ for this study reflects affordability constraints resulting from the algorithmic complexity of interface tracking.) ODT does not reproduce this gradual approach to $k^{-5/3}$ scaling with increasing $Re_{\lambda_{g}}$ because it is formulated on the basis of similarity principles that are inherently high-$Re_{\lambda_{g}}$ properties.

\section{Comparison of DNS and ODT interface statistics} \label{sec:interface}

\subsection{Cases and data-reduction procedure}
\label{sec:cases}
The comparisons of ODT and DNS results that follow are based on the flow states at $t/\tau = 0.5$. Interface structure at $t/\tau = 0.5$ (i.e., after the initially flat interface has deformed for half a large-eddy turnover time) produced by individual DNS and ODT realizations for cases 1 - 4 are shown in Fig. \ref{DNS}. Case 1 in Fig. \ref{DNS} corresponds to $W\!e_{\lambda_{g}} = \infty$ $(\sigma = 0)$, for which the interface is dynamically passive. It is apparent that  the large-scale variations of the interface are similar for each case, but smaller interfacial features are increasingly lost as $W\!e_{\lambda_{g}}$ decreases to 1.36. 

	\begin{figure}[!ht]
	\begin{center}
	\vspace{-0.5in}
		\begin{subfigure}[c]{1.1\columnwidth}
		\begin{tikzpicture}
  \node[inner sep=-3pt] (A) {\includegraphics[width=2.2in]{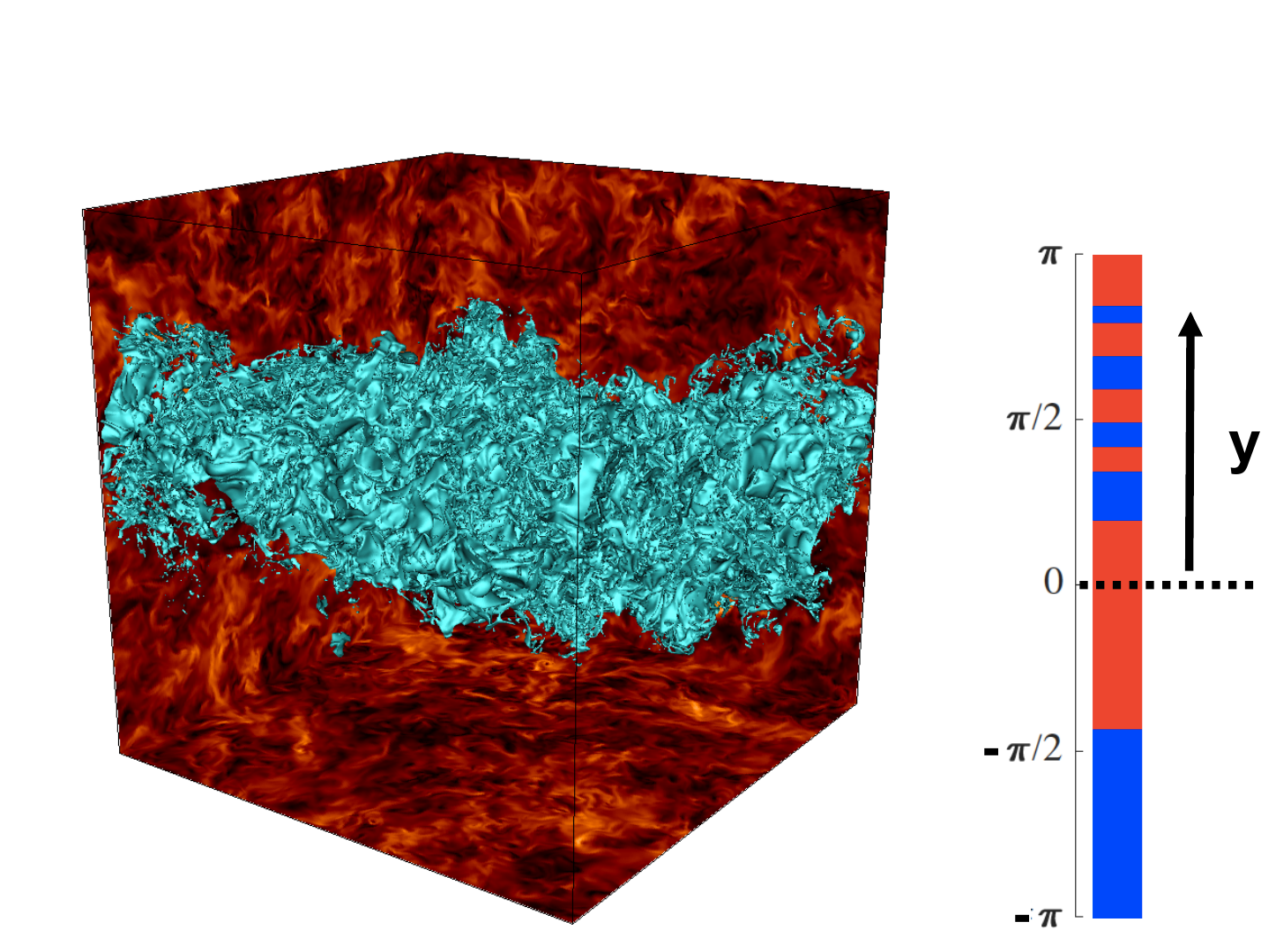}};
  \node[black] (B) at ({$(A.west)!.51!(A.east)$} |- {$(A.south)!-.06!(A.north)$}) {DNS};
  \node[black] at ({$(A.west)!.88!(A.east)$} |- {$(A.south)!-.06!(A.north)$}) {ODT};
\end{tikzpicture}
			\caption{$We_{\lambda_{g}} = \infty$}
			\label{DNSline1}
		\end{subfigure}
		\begin{subfigure}[c]{1.1\columnwidth}
		\begin{tikzpicture}
  \node[inner sep=-3pt] (A) {\includegraphics[width=2.2in]{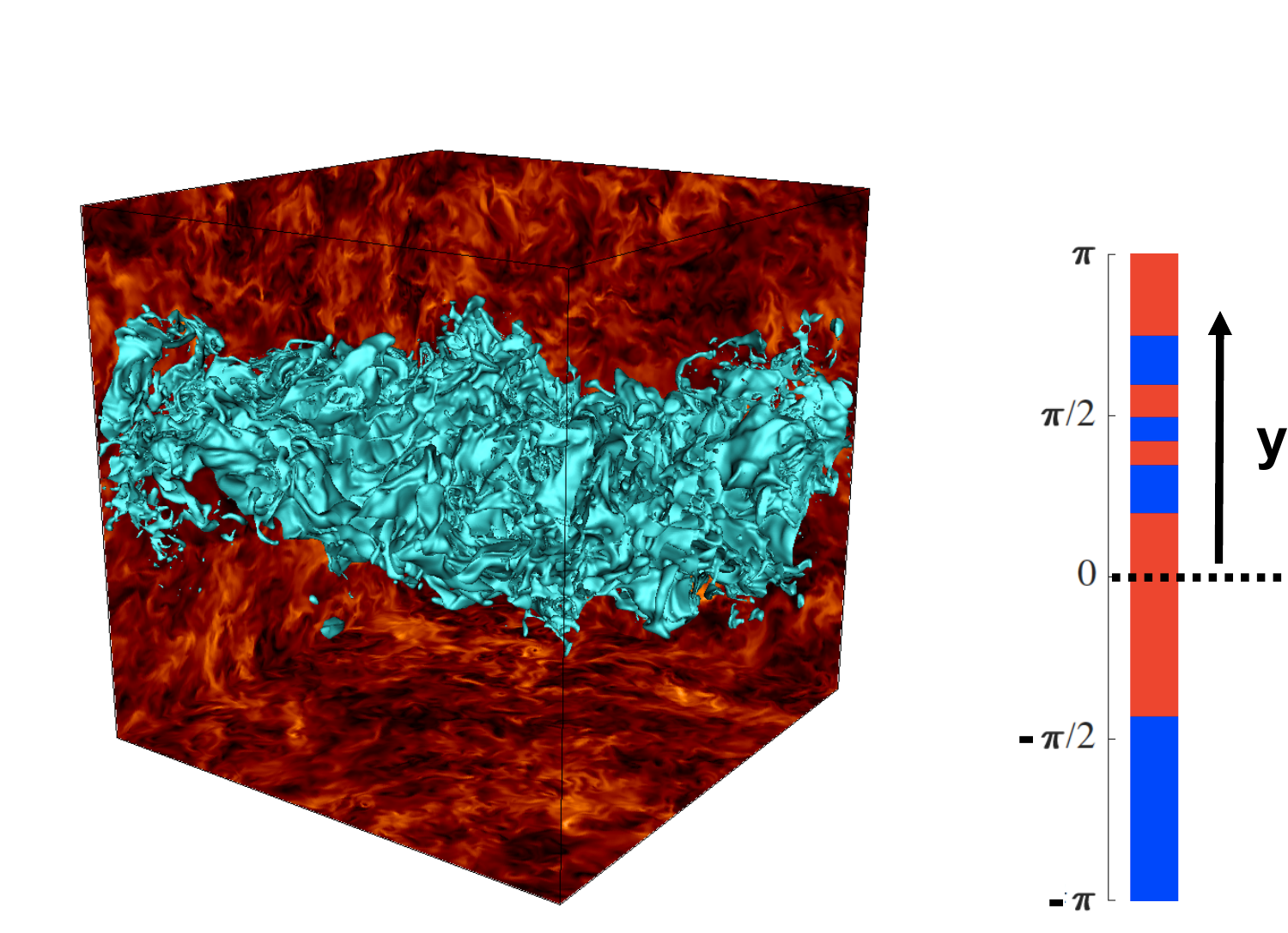}};
  \node[black] (B) at ({$(A.west)!.51!(A.east)$} |- {$(A.south)!-.06!(A.north)$}) {DNS};
  \node[black] at ({$(A.west)!.88!(A.east)$} |- {$(A.south)!-.06!(A.north)$}) {ODT};
\end{tikzpicture}
			\caption{$We_{\lambda_{g}} = 21.06$}
			\label{DNSLine1}
		\end{subfigure}
		\begin{subfigure}[c]{1.1\columnwidth}
		\begin{tikzpicture}
  \node[inner sep=-3pt] (A) {\includegraphics[width=2.2in]{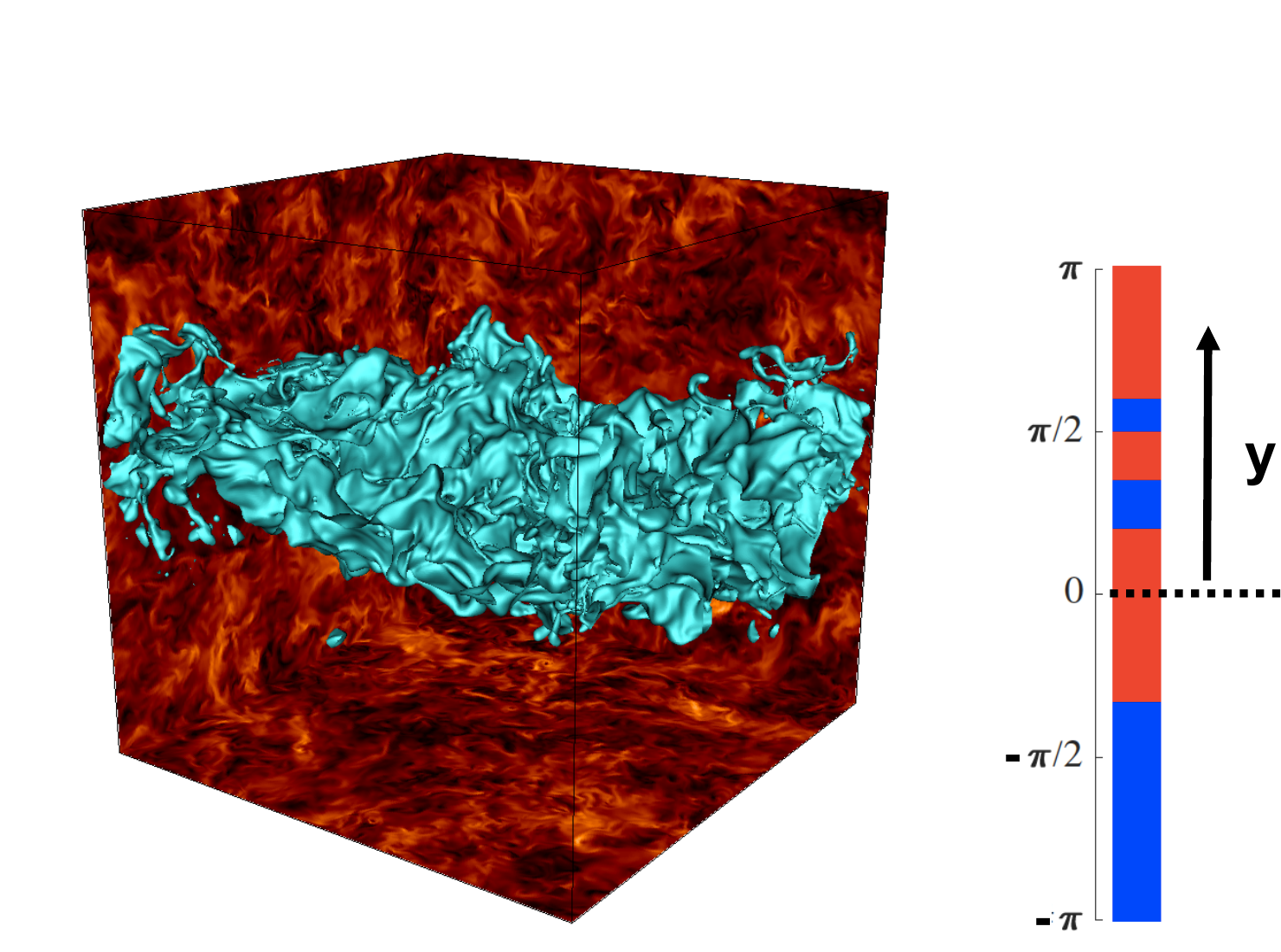}};
  \node[black] (B) at ({$(A.west)!.51!(A.east)$} |- {$(A.south)!-.06!(A.north)$}) {DNS};
  \node[black] at ({$(A.west)!.88!(A.east)$} |- {$(A.south)!-.06!(A.north)$}) {ODT};
\end{tikzpicture}
			\caption{$We_{\lambda_{g}} = 8.47$}
		\end{subfigure}
		\begin{subfigure}[c]{1.1\columnwidth}
		\begin{tikzpicture}
  \node[inner sep=-3pt] (A) {\includegraphics[width=2.2in]{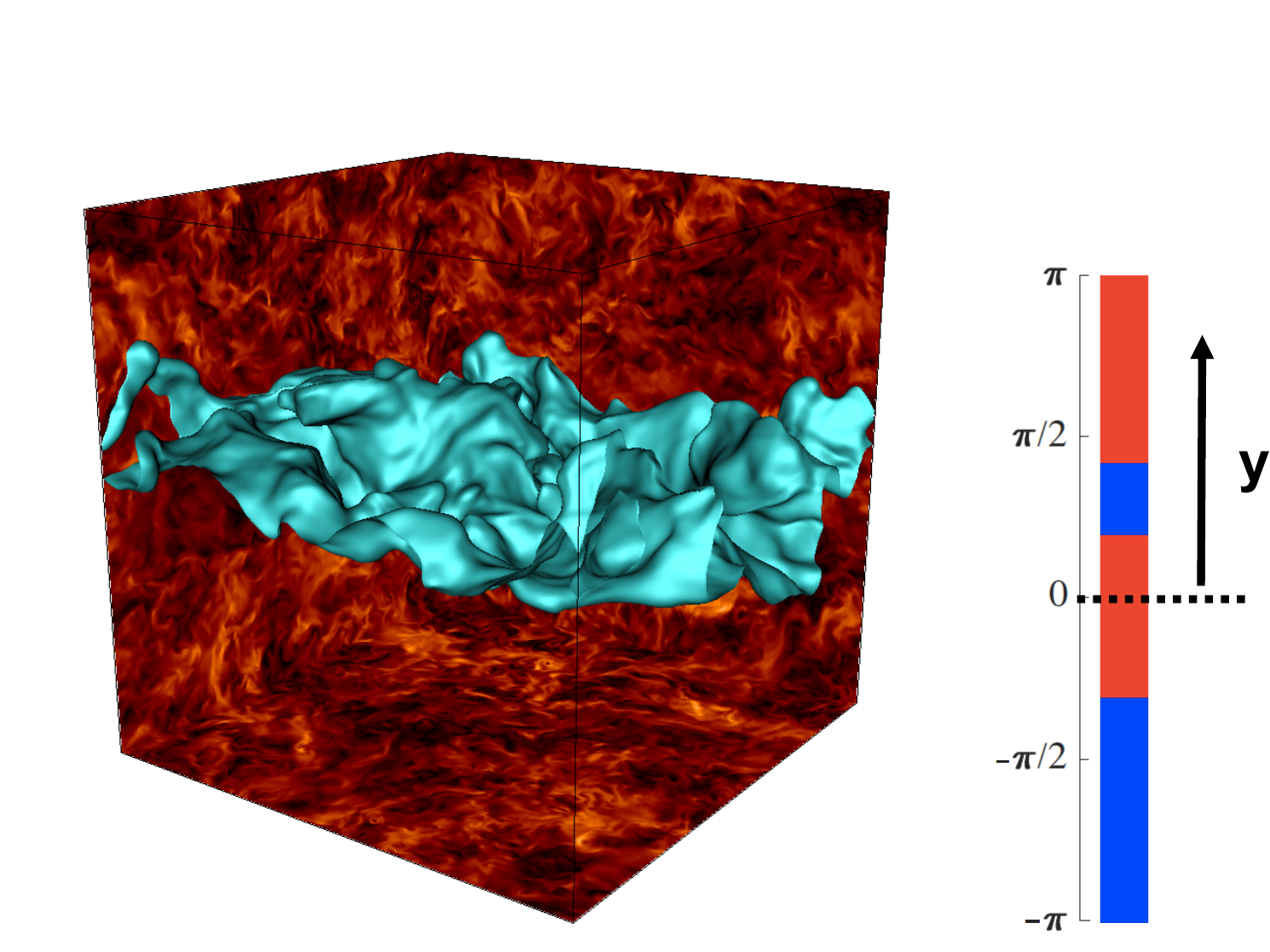}};
  \node[black] (B) at ({$(A.west)!.51!(A.east)$} |- {$(A.south)!-.06!(A.north)$}) {DNS};
  \node[black] at ({$(A.west)!.88!(A.east)$} |- {$(A.south)!-.06!(A.north)$}) {ODT};
\end{tikzpicture}
			\caption{$We_{\lambda_{g}} = 1.36$}
			\label{DNSLine2}
		\end{subfigure}
	\end{center}	
	\caption{DNS and ODT interface configurations at time $t/\tau = 0.5$. Distinct phases on the ODT line are represented by different colors. The ODT simulations use the same random number seed for all cases so that eddy-sampling fluctuations do not obscure the systematic trend.}
	\label{DNS}
\end{figure}
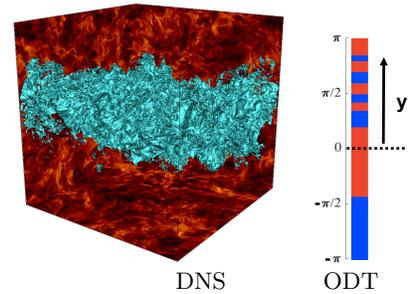
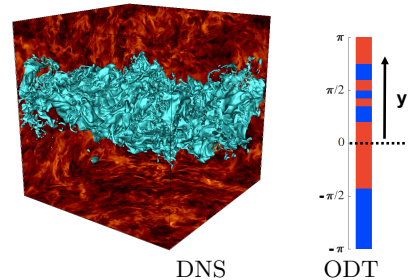
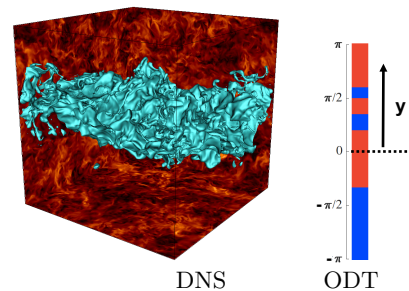
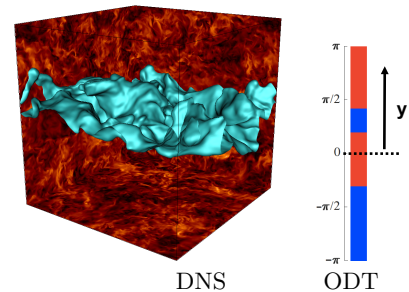

As indicated in Fig.~\ref{DNS}, the ODT line orientation ($y$ direction) is normal to the initial interface where the dashed line at $y = 0$ shows the initial interface location and the phases are distinguished by colors. Since ODT is a 1D model, the interface measure that it provides is the number of phase boundaries along the line at a given instant, where those boundaries are the model analog of intersections of a $y$-oriented line of sight with a notional interfacial surface separating the two phases. The latter information is therefore extracted from the DNS simulations for comparison to the statistics provided by ODT. This involves data collection along lines ($y$ direction) normal to the initial interface ($x$-$z$ plane), which provides statistics along 512 $\times$ 512 lines where each is analogous in terms of data analysis to an ODT domain. Although the number of such DNS lines greatly exceeds the number of ODT realizations (2000) for each case, this does not necessarily imply greater statistical precision of the DNS results because the flow states along neighboring DNS lines are highly correlated but ODT realizations are statistically independent. Because phase index labeling is arbitrary, all statistical quantities are symmetric in $y$ with respect to $y=0$, so data for positive and negative $y$ are combined and plotted over the $y$ range $[0, \pi]$. 

\subsection{Number of interfaces}
\label{NumInt}
In what follows, the term interface refers to one intersection of a $y$-oriented line of sight with the notional (in ODT) or actual (in DNS) interfacial surface. On this basis, the ensemble average value of the total number of interfaces along the line of sight at $t = 0.5 \tau$ for cases 1 - 4 is shown in Table \ref{tab:table3}.

\begin{table}
\caption{\label{tab:table3}{Mean number of interfaces in DNS and ODT simulations at at time 0.5$\tau$}}
\begin{ruledtabular}
\begin{tabular}{cccccccc}
 Case&ODT&DNS&\\
\hline
$W\!e_{\lambda_{g}} = \infty$& 9& 10& \\
$W\!e_{\lambda_{g}} = 21.06$& 4&  6&  \\
$W\!e_{\lambda_{g}}= 8.47$& 3&  5&   \\
$W\!e_{\lambda_{g}} = 1.36$& 1.6&  3&  \\
\end{tabular}
\end{ruledtabular}
\end{table}

As noted in Section \ref{Valid}, only flow-field information is used to set ODT parameters, and ODT interface results are predicted accordingly. (In Appendix \ref{sec:multi} it is explained that conversion from kinetic energy to surface-tension energy resulting from surface-area increase introduces no additional free parameters for finite $W\!e_{\lambda_{g}}$ because both energy forms are uniquely defined in ODT, albeit through simplifying assumptions.) The agreement with DNS in Table \ref{tab:table3} is good for $W\!e_{\lambda_{g}} = \infty$, with increasing under-prediction as $W\!e_{\lambda_{g}}$ decreases.

The $W\!e_{\lambda_{g}} = \infty$ result indicates that, in the absence of surface tension, ODT provides a quantitative representation of material-surface increase, which is important not only for the present application but also for other applications involving advected surfaces, such as the propagation of flames and other reacting fronts. This agreement is dependent on the choice of the parameter $C$, which scales the overall eddy rate and therefore the characteristic time for exponential increase of the number of interfaces. Specifically, the agreement indicates that the value of $C$ that gives the correct energy dissipation rate is also accurate with regard to interface development. This is not an {\it a priori} known property of the model formulation, and accordingly lends significant support to the present model application.

In this context, less accurate prediction for finite $W\!e_{\lambda_{g}}$ can be attributed in part to the ODT treatment of the abovementioned energy conversion. One assumption in that treatment is especially inaccurate at low values of $W\!e_{\lambda_{g}}$. Surface-tension energy change is the product of $\sigma$ and the surface-area change. ODT time advances surface intersections with a line of sight, from which surface-area change must be inferred. As noted in Appendix \ref{sec:multi}, this is done by assuming that the phase boundary is an isotropic random surface, consistent with the small-scale structure of high-$Re_{\lambda_{g}}$ turbulence. However, Fig.~\ref{DNS} indicates increasing anisotropy with decreasing $W\!e_{\lambda_{g}}$ for these moderate-$Re_{\lambda_{g}}$ cases.

The prefactor 2 in Eq.~(\ref{esigma}) follows directly from the assumption that the phase boundary is an isotropic random surface. If instead it is a collection of planes normal to the $y$ direction, each therefore corresponding to one point of intersection with the ODT domain, then the relation $\alpha = 2n$ in Section \ref{sec:multi} becomes instead $\alpha = n$, which eliminates the factor of 2 in Eq.~(\ref{esigma}). Figure \ref{DNS}.d suggests that the phase boundary is more plausibly idealized as primarily normal to the $y$ direction at low $W\!e_{\lambda_{g}}$ than as isotropic, so Eq.~(\ref{esigma}) potentially overstates the energy penalty for interface creation by as much as a factor of two. It is likewise plausible that this could result in half as much interface creation as a more accurate assumption about interface orientation, so this alone might explain why Table \ref{tab:table3} indicates that ODT underestimates the number interfaces by almost a factor of 2 for $W\!e_{\lambda_{g}} = 1.36$. Though an improvement might be achieved by using a more accurate case-specific prefactor in Eq.~(\ref{esigma}), such information is not available {\it a priori} when ODT is used for prediction, and the intent here is to test its predictive capability rather than to do case-specific parameter fitting. Isotropy should be a reasonably accurate assumption at higher $W\!e_{\lambda_{g}}$ and $Re_{\lambda_{g}}$ values of practical interest, so the agreement should improve relative to the finite-$W\!e_{\lambda_{g}}$ results in Table \ref{tab:table3}.

Figure \ref{DNS} also indicates that the surfaces are smoother at lower $W\!e_{\lambda_{g}}$, implying that the resistance of the surfaces to wrinkling by small eddies might suppress nearby eddy motions, much as a turbulent boundary layer becomes increasingly laminar as the wall is approached. Then the creation of surface overhangs, which is required to produce multiple intersections with a $y$-directed line of sight, might occur largely due to non-vortical shearing motions applied to locally tilted surface elements. ODT contains no representation of non-vortical motion, so it does not capture this mechanism and therefore would not reproduce the full extent of interface generation for conditions under which this mechanism is important.

To put these observations in context, the present comparisons are constrained by the limited parameter space that is accessible using DNS, which does not include the high-$Re_{\lambda_{g}}$ regime that ODT is designed to represent most accurately. Additionally, order-unity errors are substantial in the context of point predictions, but ODT comparisons to jet breakup, which likewise show errors of this magnitude, capture trends extending over orders of magnitude in Reynolds and Weber number, a situation in which order-one point-prediction errors are inconsequential \cite{movaghar2017numerical}. In this context, the contributions of the present study are twofold: the simple flow configuration allows a simpler, less empirical ODT formulation than in \cite{movaghar2017numerical} to be used, and detailed statistics that DNS can provide allow comparison of structural features of the flow as well as global properties such as the data in Table \ref{tab:table3}. These structural features are explored next.

\subsection{Interface number density}
\label{NumDen}

The interface number density, shown in Fig.~\ref{Density}, is obtained by taking the $y$ derivative of the total number of interfaces between $ y $ and $ - y$ and dividing the result by $2y$.  Integration of the number density over $[0, \pi]$ gives the ensemble average value of the total number of interfaces on one side of the initial interface, which is half of the number of interfaces on the whole domain $[-\pi, \pi]$, where the latter is the quantity shown in Table \ref{tab:table3}. This procedure yields the unshifted profiles in Fig.~\ref{Density}, which shows that the ODT profiles are substantially narrower than the DNS profiles. 

The DNS images in Fig.~\ref{DNS} suggest that the surface is subject to large-scale displacements in the $y$ direction superimposed on smaller-scale vortically induced displacements. Large-scale displacements are facilitated by the periodic boundary conditions, which in principle allow non-vortical streaming motions in the positive and negative $y$ direction.

Adopting the hypothesis that such motions, which ODT cannot capture, contribute to the greater broadening of the DNS profiles relative to the ODT profiles, the data is re-processed to eliminate the possible effects of this mechanism. The number of interfaces along a line of sight at a given instant must be odd because the pure-phase regions beyond the mixed-phase zone have opposite phase, so there must be an odd number of phase flips along any trajectory that extends through the mixed-phase zone. Then there is one `median' interface such that it has an equal number of interfaces on either side of its location. The displacement of the median interface relative to $y=0$ is deemed to be a measure of the displacement of the mixed-phase zone due to large-scale $y$-oriented motions. Therefore each instantaneous state along the line of sight is shifted so that the median interface is relocated to $y=0$, as illustrated in Fig.~\ref{Median}, and this modified data ensemble is then used to obtain `shifted' interface number-density profiles. The median interface is excluded from this data reduction because its displacement to $y=0$ implies infinite number density at $y=0$ (a $\delta$-function spike) that reflects data conditioning rather than a physical effect.

ODT interfaces are not subject to non-vortical displacements so the ODT profiles in Fig.~\ref{Density} that are based on shifted data are not much different from those that are based unshifted data, but the effect of the shifts on the DNS profiles is pronounced and brings them into closer conformance to the ODT results. Full agreement of ODT and DNS profiles is impossible because of the differences between the areas under each pair curves that is implied by the results in Table \ref{tab:table3}. Given this unavoidable discrepancy, the agreement of the shapes of the profiles is noteworthy, especially the low-$y$ plateau-cliff structure that is seen only for $W\!e_{\lambda_{g}} = 1.36$. The agreement of profile shapes seen in the shifted results suggests that ODT is at least somewhat representative of the structure of the DNS mixed-phase zone in a Lagrangian sense, meaning in a reference frame in which the ODT and DNS mixed-phase zones roughly coincide. Accordingly, this Lagrangian interpretation of the present ODT formulation is adopted, so results that follow are based on shifted data. 

\begin{figure}[!ht]
\begin{center}
\begin{tikzpicture}
  \node[inner sep=-3pt] (A) {\includegraphics[width=2in]{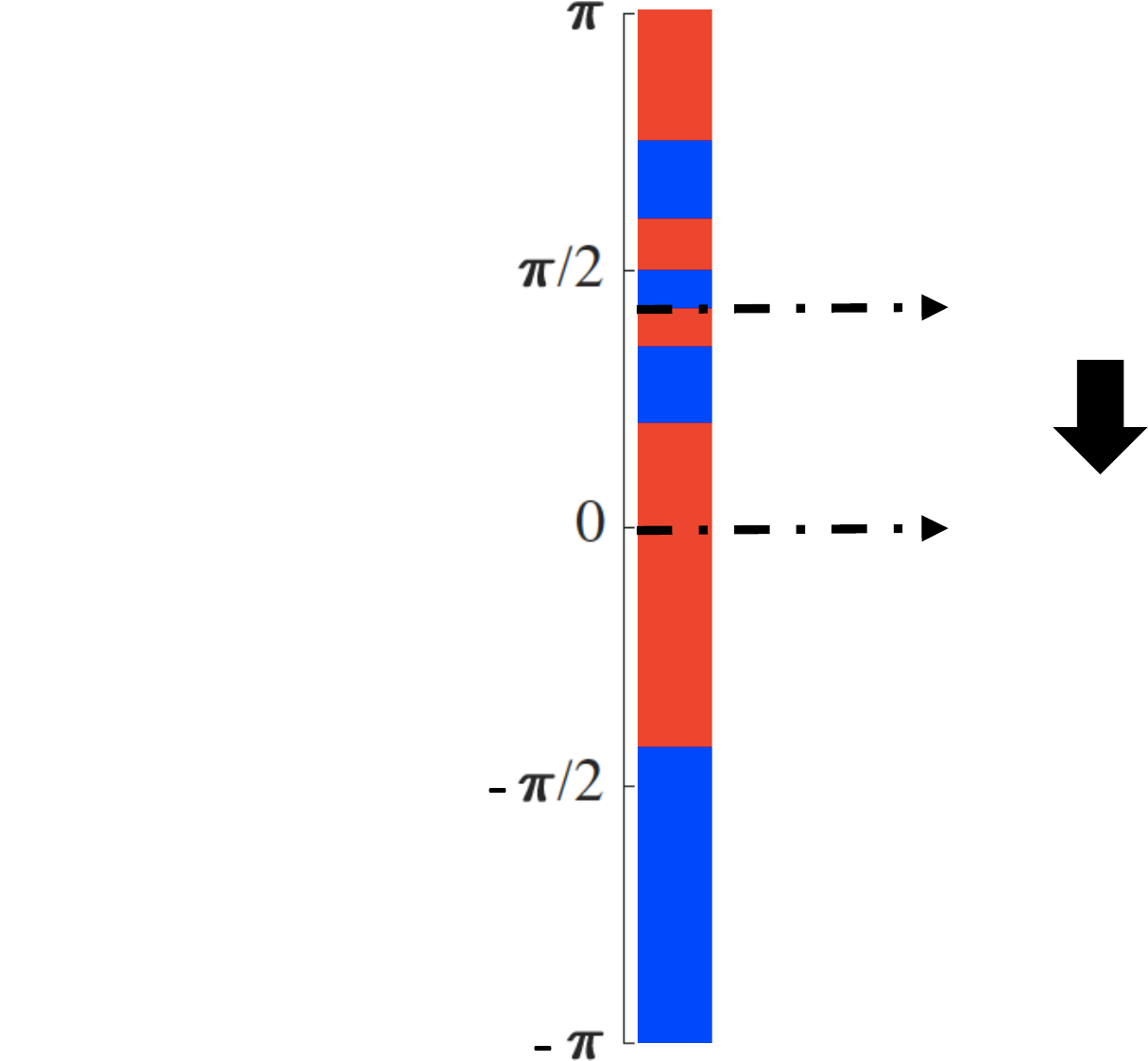}};
  \node[black] (B) at ({$(A.west)!.98!(A.east)$} |- {$(A.south)!.4!(A.north)$}) {Origin};
  \node[black] at ({$(A.west)!1.04!(A.east)$} |- {$(A.south)!.8!(A.north)$}) {Median interface location};
\end{tikzpicture}
\end{center}
\caption{Schematic of the shift of the location of the median interface to the origin.} 
\label{Median} 
\end{figure}

\begin{figure}[!ht]
	\begin{center}
\begin{subfigure}[c]{0.55\columnwidth}
		\hspace*{-1.2em}
		\begin{tikzpicture}
  \node[inner sep=0pt] (A){\includegraphics[width=\textwidth]{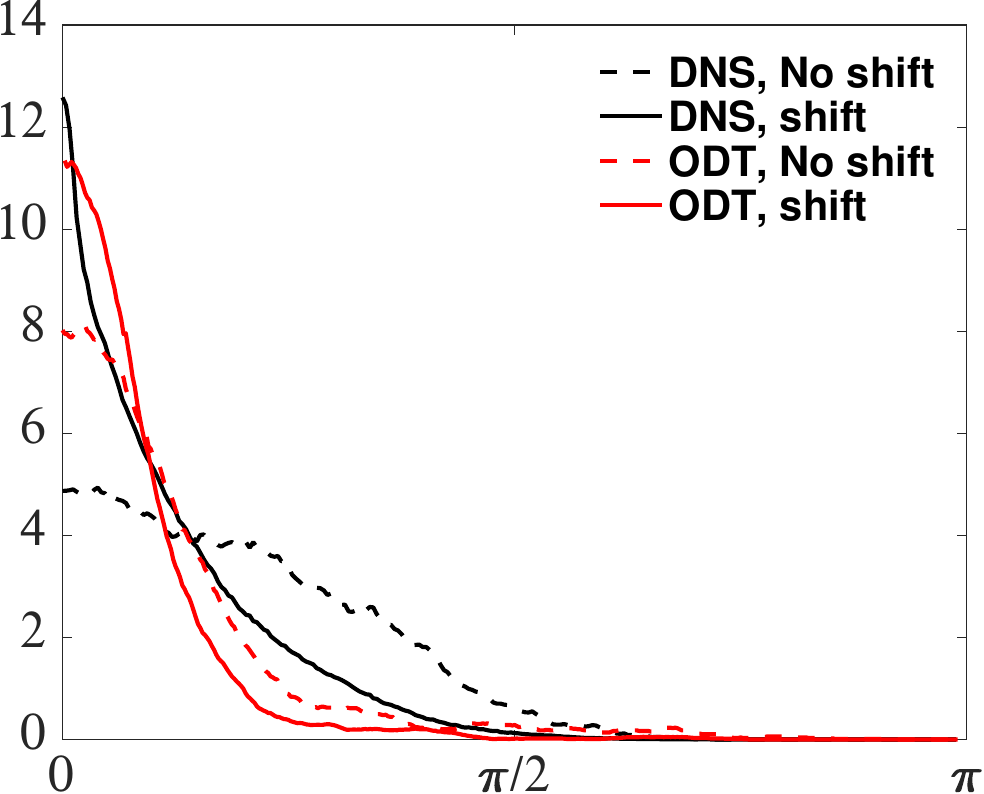}};
  \node[black] (B) at ($(A.south)!-.07!(A.north)$) {y};
  \node[black,rotate=90] (C) at ($(A.west)!-.07!(A.east)$) {Interface number density};
\end{tikzpicture}
			\caption{$We_{\lambda_{g}} = \infty$}
			\label{PhaseInf}
		\end{subfigure}
		\begin{subfigure}[c]{0.55\columnwidth}
		\hspace*{-1.2em}
		\begin{tikzpicture}
  \node[inner sep=0pt] (A) {\includegraphics[width=\textwidth]{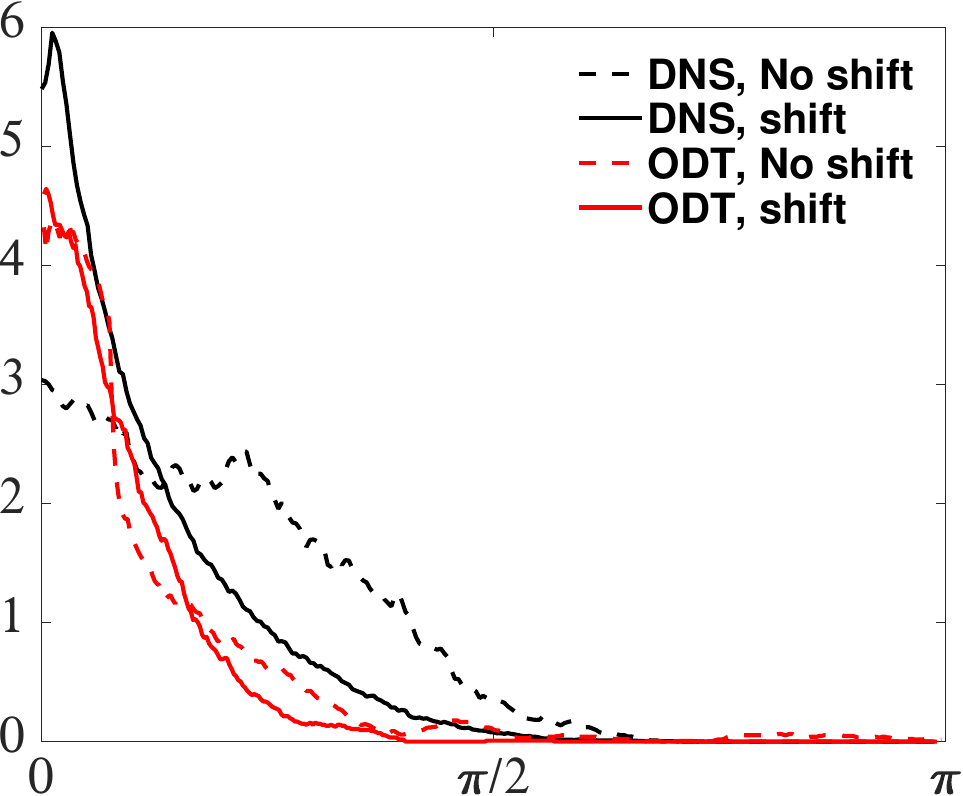}};
  \node[black] (B) at ($(A.south)!-.07!(A.north)$) {y};
  \node[black,rotate=90] (C) at ($(A.west)!-.07!(A.east)$) {Interface number density};
\end{tikzpicture}
			\caption{$We_{\lambda_{g}} = 21.06$}
		\end{subfigure}
		\begin{subfigure}[c]{0.6\columnwidth}
		\hspace*{-1.2em}
\begin{tikzpicture}
  \node[inner sep=0pt] (A) {\includegraphics[width=\textwidth]{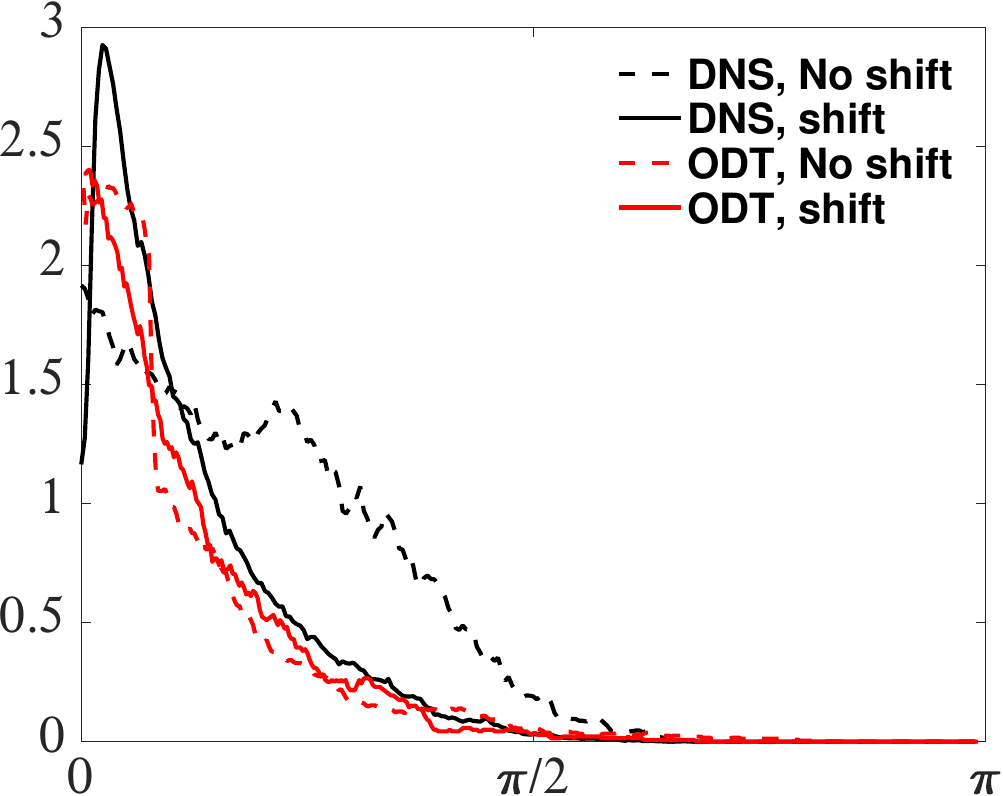}};
  \node[black] (B) at ($(A.south)!-.07!(A.north)$) {y};
  \node[black,rotate=90] (C) at ($(A.west)!-.07!(A.east)$) {Interface number density};
\end{tikzpicture}
			\caption{$We_{\lambda_{g}} = 8.47$}
		\end{subfigure}
		\begin{subfigure}[c]{0.6\columnwidth}
		\hspace*{-1.2em}
		\begin{tikzpicture}
  \node[inner sep=0pt] (A) {\includegraphics[width=\textwidth]{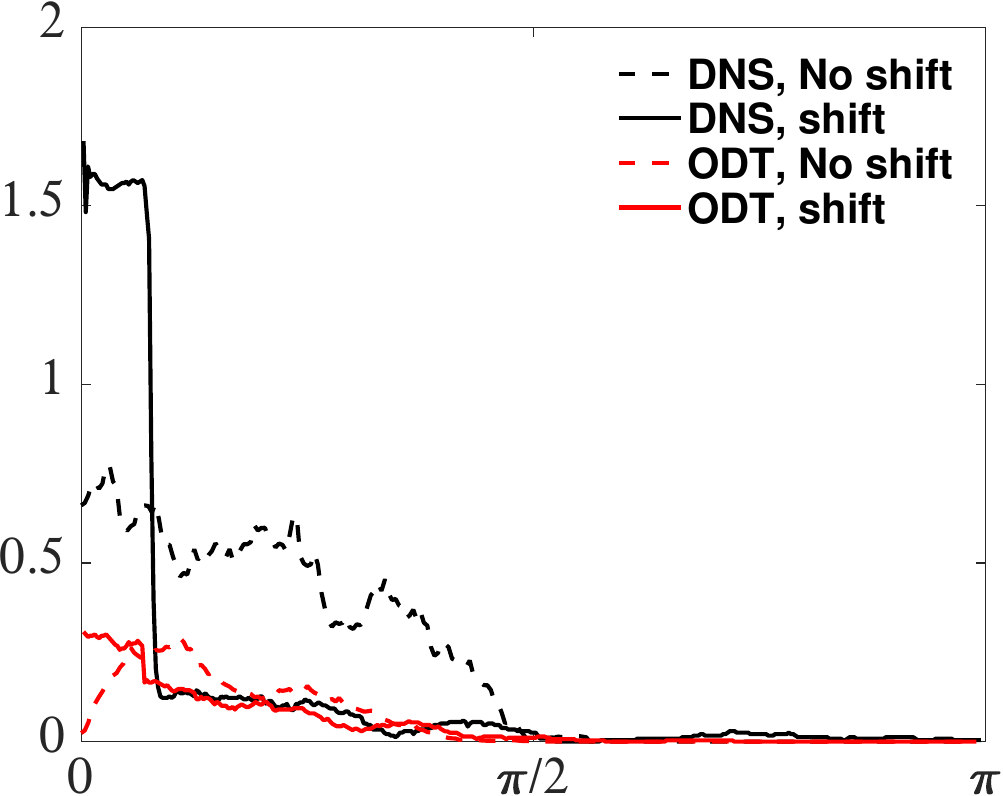}};
  \node[black] (B) at ($(A.south)!-.07!(A.north)$) {y};
  \node[black,rotate=90] (C) at ($(A.west)!-.07!(A.east)$) {Interface number density};
\end{tikzpicture}
			\caption{$We_{\lambda_{g}} = 1.36$}
			\label{DensityInf}
		\end{subfigure}	
	\end{center}	
	\caption{DNS and ODT interface number density at time $t/\tau = 0.5$.}
	\label{Density}
\end{figure}
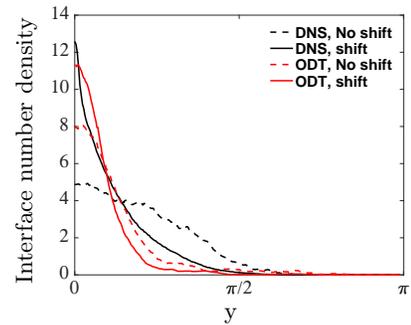
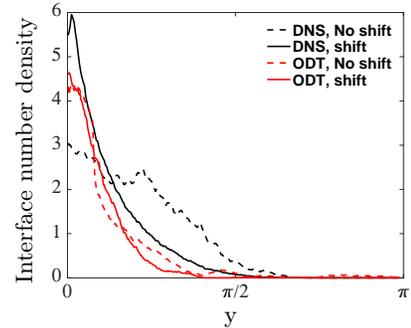
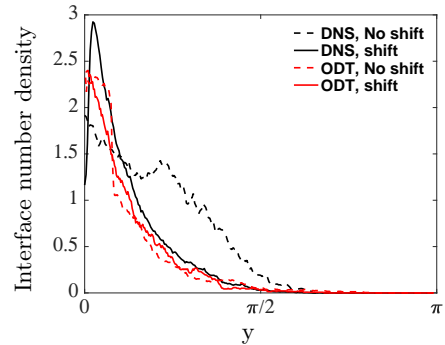
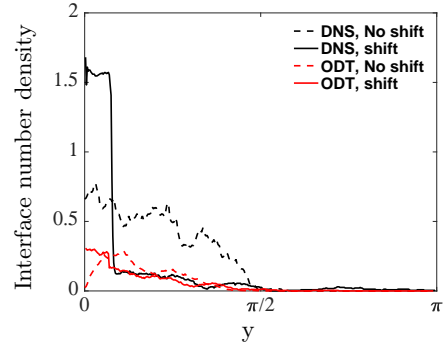

\subsection{Same-phase probability} \label{sec:same}

Two-point statistics provide another perspective on interface structure. Power spectra are applicable in homogeneous directions, but the phase index is statistically homogeneous only in $x$ and $z$, while the ODT domain is oriented in the inhomogeneous direction $y$. Since the power spectrum of a zero-mean property $p$ is the Fourier transform of its two-point covariance $ R(y_1, y_2) = \langle p(y_1) p(y_2) \rangle$, the latter embodies the same information and can therefore be used. In homogeneous directions, the covariance depends only on  $|y_1 - y_2|$, but otherwise it is irreducibly dependent on both arguments.

Taking $p$ to be the phase index, with possible values $\pm 1$, then $p(y_1) p(y_2)$ is $+1$ ($-1$) if the phases at the two locations are the same (different). Then $[ 1+ p(y_1) p(y_2)] / 2$ is 1 or 0 in the respective instances, so 
$S = (1 + R) / 2$ is the probability of finding the same phase at the two locations. Since this embodies the same information as $R$, the same-phase probability $S(y_1, y_2)$ is used as a representative measure of the two-point structure of the interface.

This statistic has been evaluated using simulation data, and it is found that the features of interest are largely captured by the results for $y_2 = -y_1$. Adopting this specialization, $S(y)$ is defined as the probability that the same phase index is found at locations $y$ and $-y$ relative to a specified origin. Based on the results in Section \ref{NumDen}, the origin is taken to be the location of the median interface along the line of sight. In Fig.~\ref{SamePhaseShift}, the same-phase probability is plotted as a function of the absolute distance $\Delta y$ between $y_1$ and $y_2$, so the plotted function is $S(\Delta y / 2)$.

Due to the coordinate shift, the median interface is located at the origin, so $S(0)$ is strictly speaking undefined, but because it reduces to a single-point statistic at $y=0$, it is deemed to be unity. For a vanishingly small but nonzero argument, there is vanishing likelihood of more than one interface in $[-y, y]$, so $S(y)$ converges to zero in this limit and therefore, as defined, is discontinuous at $y=0$, which is immaterial because $S(0)$ is a non-informative quantity.

The time $t = 0.5 \tau$ was chosen for presentation of results because it is a time when the mixed-phase zone occupies an order-one fraction of the computational domain but does not closely approach the domain boundaries. Accordingly, for values of the argument of $S$ corresponding to locations near the boundaries, the phase indices remain at their initial values and therefore are never the same, giving $S = 0$, as seen in the plots.

The $y$ location of the peak of $S$ is a signature of the typical distance from the median interface to its nearest neighbor, thus identifying an interface-separation microscale. As expected, both DNS and ODT indicate that this microscale increases with decreasing $W\!e_{\lambda_{g}}$. ODT predicts the location of the peak accurately except for the lowest $W\!e_{\lambda_{g}}$ value, for which both DNS and ODT indicate a broad peak, making the location of the peak inherently hard to predict. In Section \ref{sec:crit}, a precisely defined microscale is used to compare ODT results to theory as well as DNS.

A high peak value of $S$ corresponds to low variability of the spatial distribution of adjacent interfaces, and vice versa. As $W\!e_{\lambda_{g}}$ decreases, fewer interfaces intersect the 1D line of sight, consistent with greater spatial variability and hence broader profile peaks for lower $W\!e_{\lambda_{g}}$. Similarly, Table \ref{tab:table3} indicates fewer ODT interfaces than DNS interfaces, consistent with ODT peak heights that are lower than DNS peak heights in Fig.~\ref{SamePhaseShift}.

Overall, the same-phase probability is found to encode significant structural information. ODT is seen to capture the main $W\!e_{\lambda_{g}}$ dependencies with a degree of accuracy that is commensurate with the discrepancies indicated in Table \ref{tab:table3}.


\begin{figure}[!ht]
	\begin{center}
		\begin{subfigure}[c]{0.55\columnwidth}
		\hspace*{-1.2em}
					\begin{tikzpicture}
  \node[inner sep=0pt] (A) {\includegraphics[width=\textwidth]{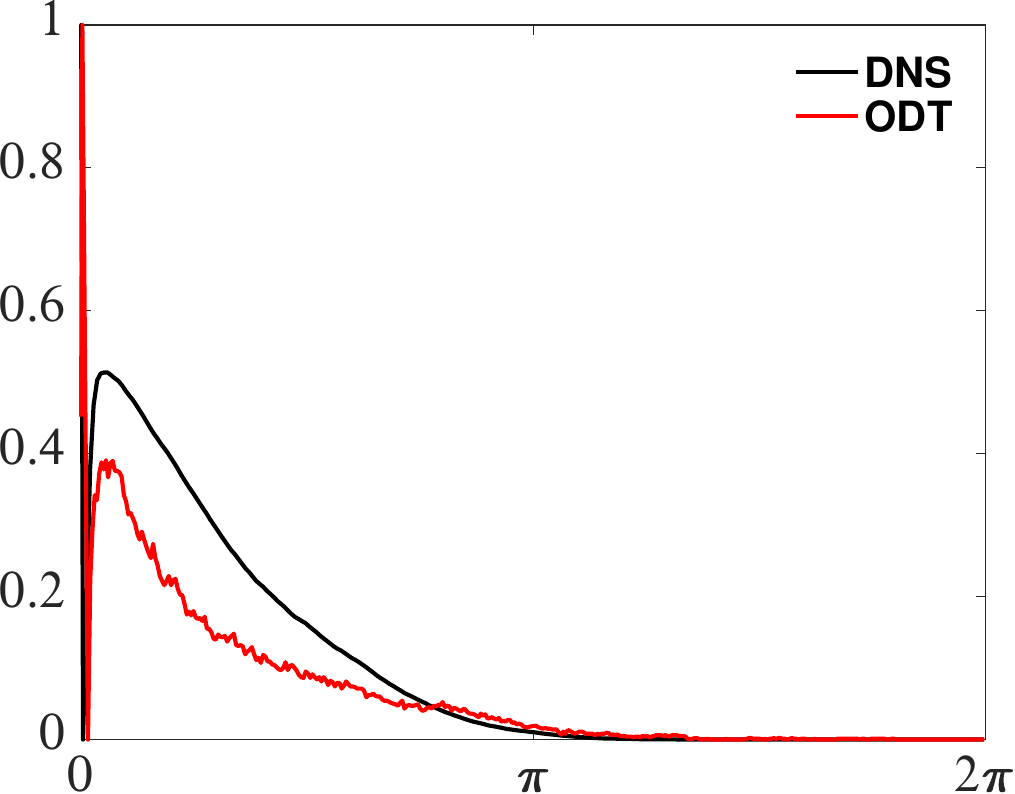}};
  \node[black] (B) at ($(A.south)!-.07!(A.north)$) {$\Delta y$};
  \node[black,rotate=90] (C) at ($(A.west)!-.07!(A.east)$) {Same phase probability};
\end{tikzpicture}
			\caption{$W\!e_{\lambda_{g}} = \infty$}
			\label{PhaseShift1}
		\end{subfigure}
		\begin{subfigure}[c]{0.55\columnwidth}
		\hspace*{-1.2em}
		\begin{tikzpicture}
  \node[inner sep=0pt] (A) {\includegraphics[width=\textwidth]{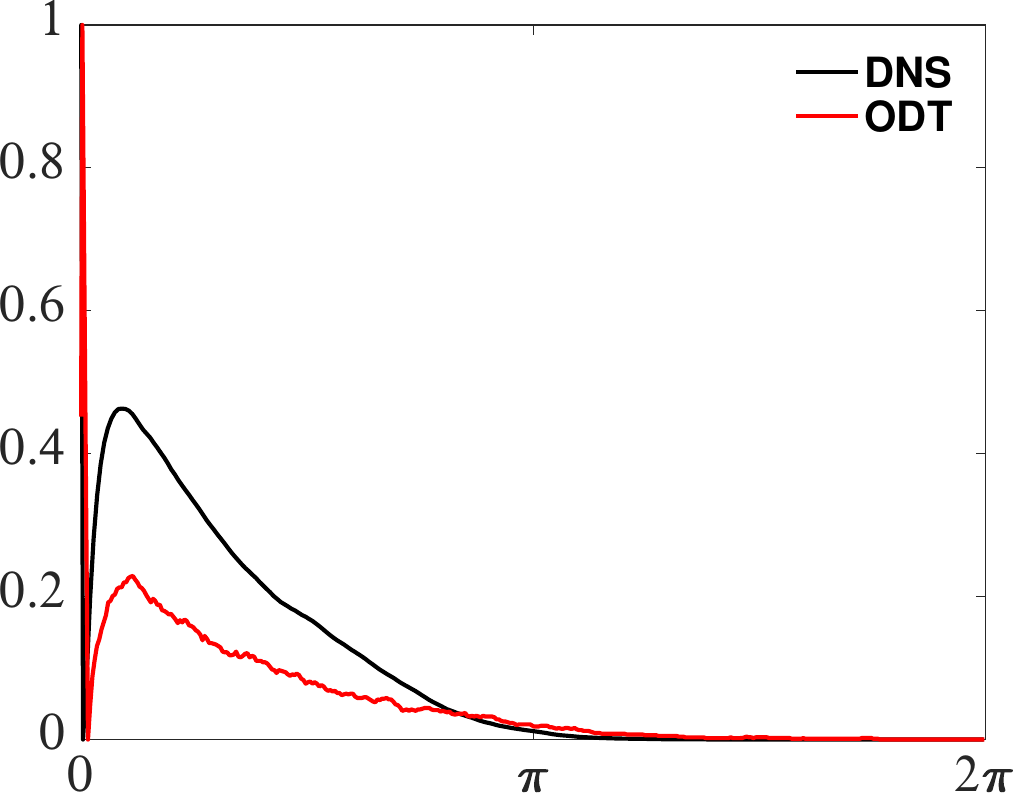}};
  \node[black] (B) at ($(A.south)!-.07!(A.north)$) {$\Delta y$};
  \node[black,rotate=90] (C) at ($(A.west)!-.07!(A.east)$) {Same phase probability};
\end{tikzpicture}
			\caption{$W\!e_{\lambda_{g}} = 21.06$}
			\label{PhaseShift2}
		\end{subfigure}
		\begin{subfigure}[c]{0.55\columnwidth}
		\hspace*{-1.2em}
		\begin{tikzpicture}
  \node[inner sep=0pt] (A) {\includegraphics[width=\textwidth]{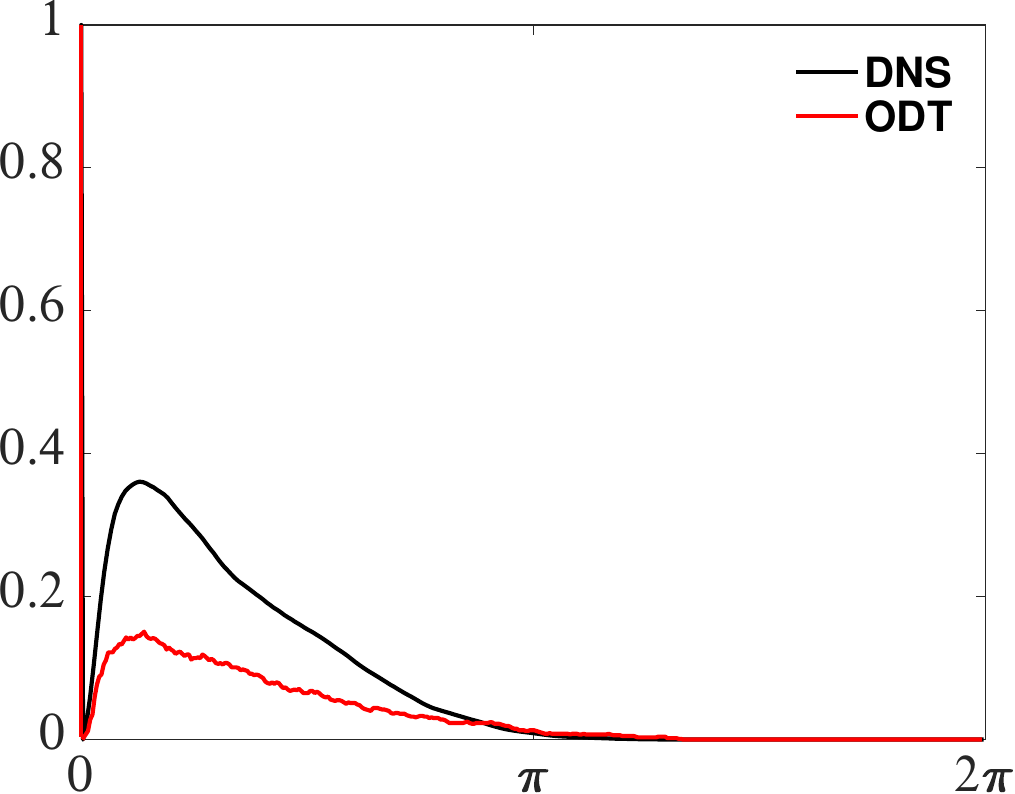}};
  \node[black] (B) at ($(A.south)!-.07!(A.north)$) {$\Delta y$};
  \node[black,rotate=90] (C) at ($(A.west)!-.07!(A.east)$) {Same phase probability};
\end{tikzpicture}
			\caption{$W\!e_{\lambda_{g}} = 8.47$}
			\label{PhaseShift3}
		\end{subfigure}
		\begin{subfigure}[c]{0.55\columnwidth}
		\hspace*{-1.2em}
		\begin{tikzpicture}
  \node[inner sep=0pt] (A) {\includegraphics[width=\textwidth]{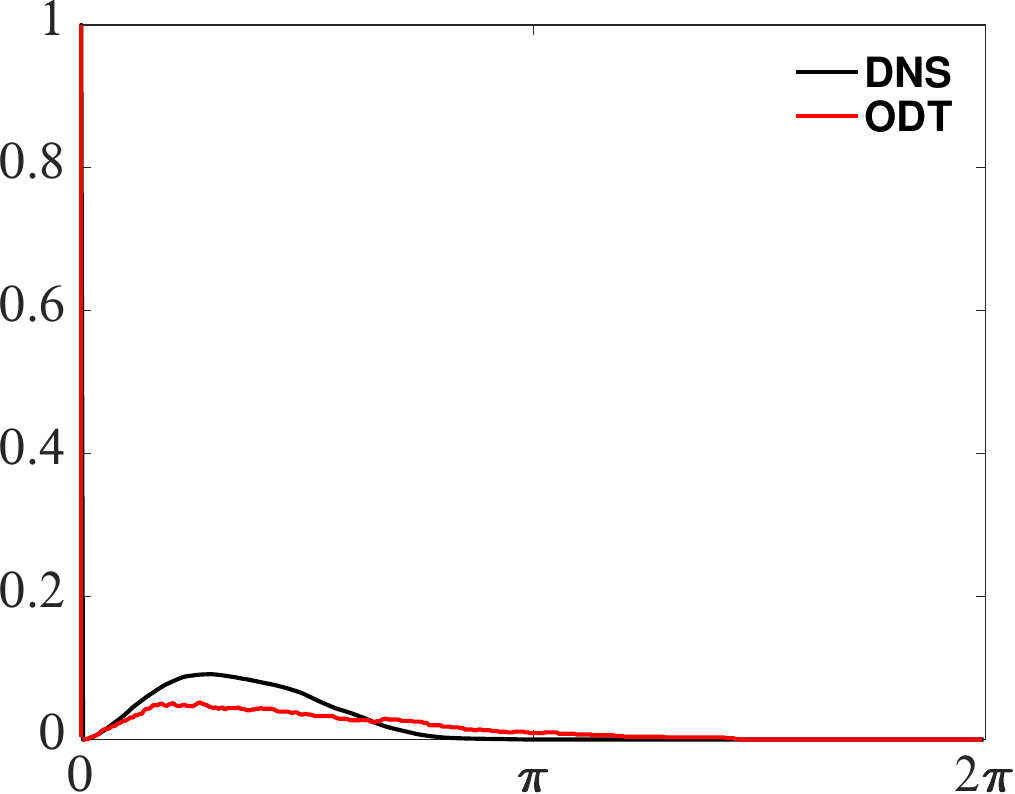}};
  \node[black] (B) at ($(A.south)!-.07!(A.north)$) {$\Delta y$};
  \node[black,rotate=90] (C) at ($(A.west)!-.07!(A.east)$) {Same phase probability};
\end{tikzpicture}
			\caption{$W\!e_{\lambda_{g}} = 1.36$}
			\label{phaseShift4}
		\end{subfigure}
	\end{center}	
	\caption{Same phase probability statistics for DNS and ODT at time $t/\tau = 0.5$.}
	\label{SamePhaseShift}
\end{figure}
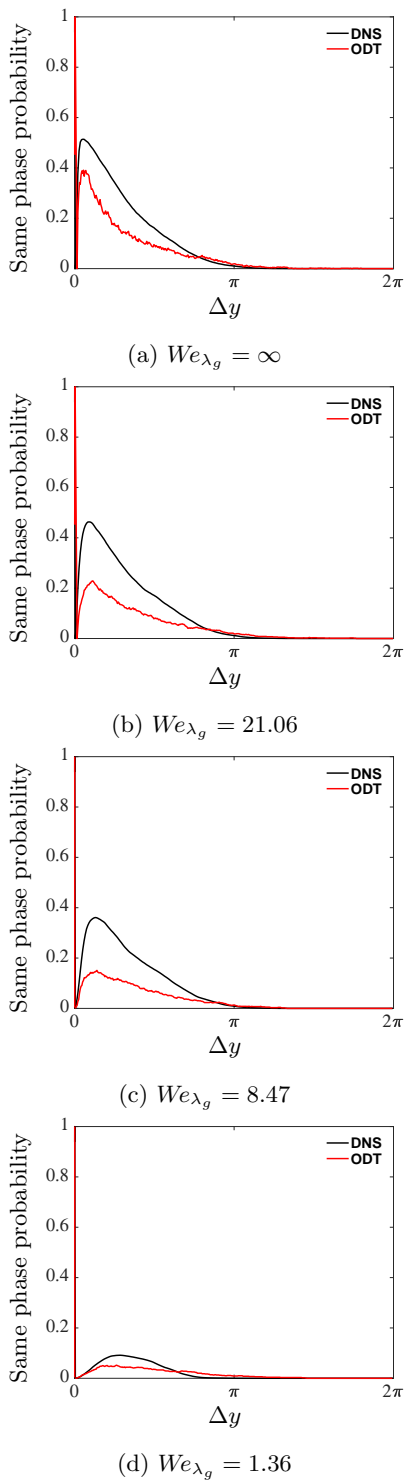

\section{Sensitivity to mesh resolution} \label{sec:mesh}

A practical feature of the ODT surface-tension formulation is that it might be useful as a subgrid closure for large-eddy simulation of turbulence that includes, e.g., an under-resolved volume-of-fluids (VOF) treatment of phase-interface evolution. This requires the model to be run at $W\!e_{\lambda_{g}}$ and $Re_{\lambda_{g}}$ values corresponding to the VOF cell Weber and Reynolds numbers. To identify the ODT resolution requirements for these conditions, $W\!e_{\lambda_{g}} = 100$ and $Re_{\lambda_{g}} = 500$ are chosen as representative values. This is addressed within a more general investigation of the numerical requirements for ODT application in the parameter space beyond the DNS-accessible regime.

ODT is numerically implemented using a specially designed adaptive mesh that does not limit the resolution of the spacing of interfaces along the line of sight. In contrast, the DNS is implemented using a uniform grid spacing. The present study compares ODT predictions to DNS results, so a minimum resolution $\Delta y_{DNS} = 2 \pi / 512$ was superimposed on the adaptive mesh algorithm to enforce the same resolution in ODT as in DNS. This was done by forcing adaptive-mesh cells to merge as needed so that no cell size in ODT simulation falls below the DNS cell size. Eddies are required to overlap a minimum number of ODT mesh cells, so this can cause suppression of eddies that might otherwise be implemented, although this does not happen if the lower bound on cell size is sufficiently small.

As implied by Fig.~\ref{DNS}, the formation of interfacial corrugation is suppressed on length scales smaller than a reference length scale that depends on Weber number. This dependence and dependence on Reynolds number are examined in detail in Section \ref{sec:crit}. Before this is done, sensitivity to mesh resolution is investigated computationally.

For the case $W\!e_{\lambda_{g}} = 21.06$ and $Re_{\lambda_{g}} = 155$, Fig. \ref{MeshStudy21} shows that the DNS mesh resolution is sufficient to capture the total number of interfaces in ODT simulations. For $W\!e_{\lambda_{g}} = 100$ and $Re_{\lambda_{g}} = 155$, the ODT results shown in Fig.~\ref{MeshStudy100} indicate that a resolution of $0.1 \Delta y_{DNS}$ is sufficient with respect to this measure of convergence, while the choice $\Delta y_{DNS}$ gives increasingly inaccurate results as the flow evolves. Further refinement of the DNS resolution would be very costly, so the contrast between Figs.~\ref{MeshStudy21} and \ref{MeshStudy100} indicates that extension of the parameter space of the DNS runs is impractical.

\begin{figure}[ht!]
        \centering     
        \begin{tikzpicture}
  \node[inner sep=0pt] (A) {\includegraphics[width=0.35\textwidth]{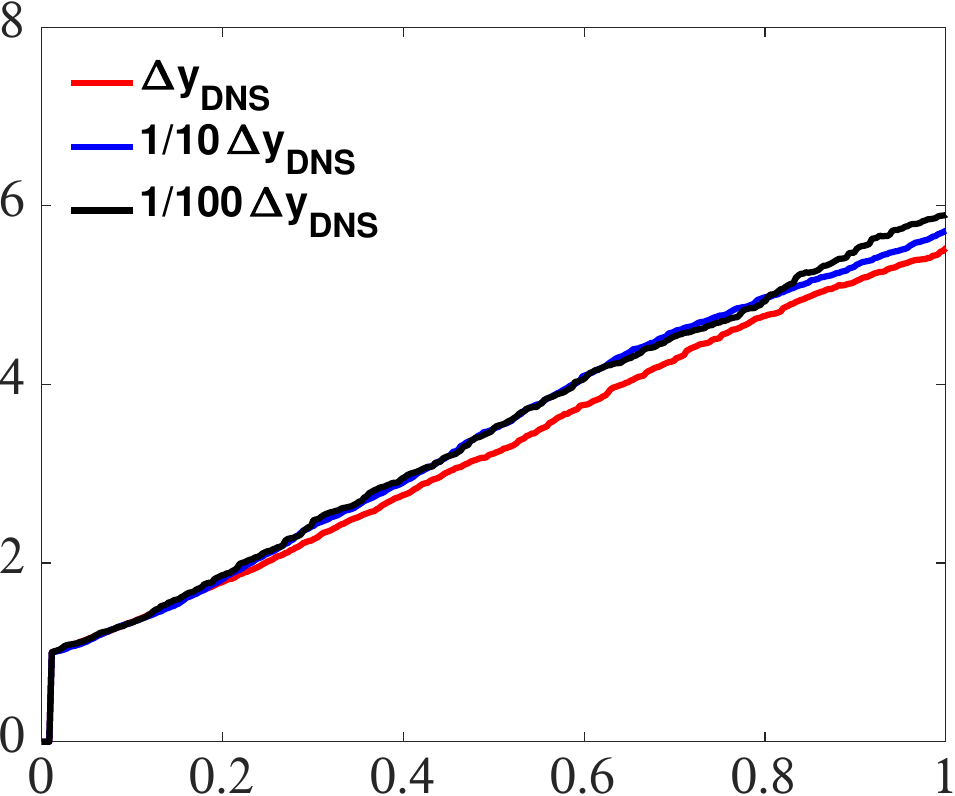}};
  \node[black] (B) at ($(A.south)!-.07!(A.north)$) {$t/\tau$};
  \node[black,rotate=90] (C) at ($(A.west)!-.07!(A.east)$) {Mean number of interfaces};
\end{tikzpicture}   
\caption{Mesh resolution sensitivity of the mean number of interfaces predicted by ODT for $W\!e_{\lambda_{g}} = 21.06$ and $Re_{\lambda_{g}} = 155$. }
\centering
\label{MeshStudy21}
\end{figure}

\begin{figure}[ht!]
        \centering
  \begin{tikzpicture}
  \node[inner sep=0pt] (A) {\includegraphics[width=0.35\textwidth]{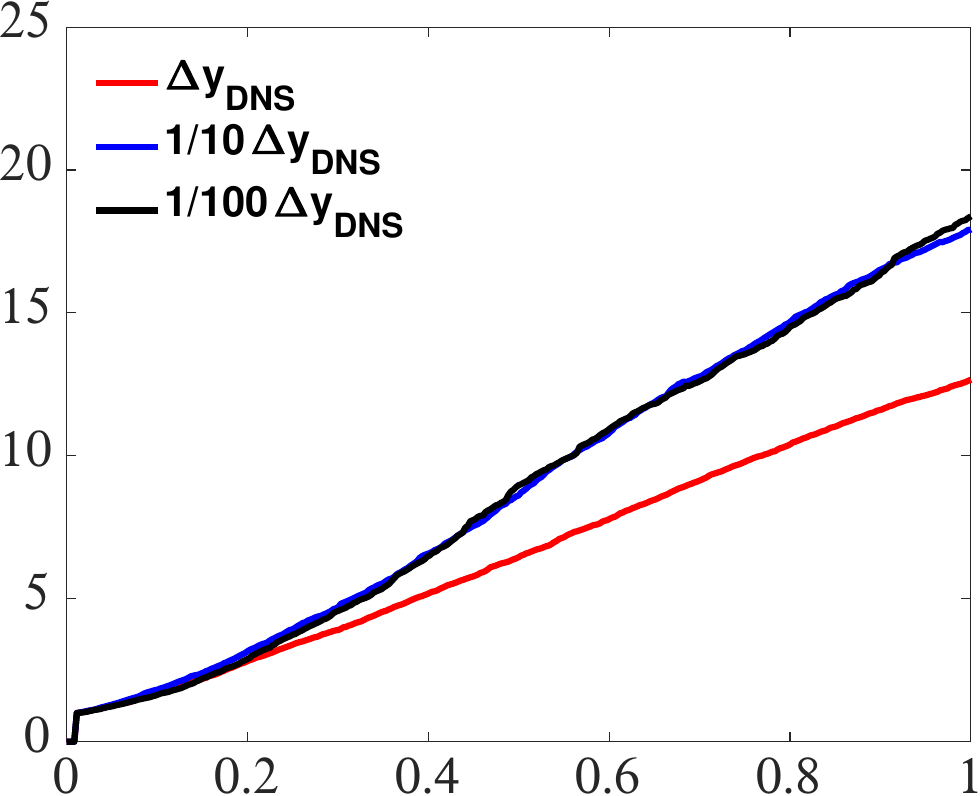}};
  \node[black] (B) at ($(A.south)!-.07!(A.north)$) {$t/\tau$};
  \node[black,rotate=90] (C) at ($(A.west)!-.07!(A.east)$) {Mean number of interfaces};
\end{tikzpicture}      
\caption{Mesh resolution sensitivity of the mean number of interfaces predicted by ODT for $W\!e_{\lambda_{g}} = 100$ and $Re_{\lambda_{g}} = 155$.}
\centering
\label{MeshStudy100}
\end{figure}

Turning now to $Re_{\lambda_{g}} = 500$, the first consideration is the ODT energy spectrum for the HIT ($W\!e_{\lambda_{g}} = \infty$) case. As in Section \ref{Valid}, the HIT simulation was initialized so that the flow state relaxed to HIT structure before reaching the target value of $Re_{\lambda_{g}}$, which in this case is 500. The one-dimensional energy spectra at the time of interface insertion are shown for various mesh resolutions in Fig.~\ref{spectrum500}. It is seen that a resolution of $0.1 \Delta y_{DNS}$ is sufficient to resolve the spectrum. For the Weber-number target $W\!e_{\lambda_{g}} = 100$, Fig.~\ref{MeshStudy500} indicates that higher resolution is needed to capture all the interfaces, showing that interface resolution is the most stringent requirement for this case as well as for the case shown in Fig.~\ref{MeshStudy100}.

\begin{figure}[ht!]
\begin{center}
\begin{tikzpicture}
  \node[inner sep=0pt] (A) {\includegraphics[width=0.4\textwidth]{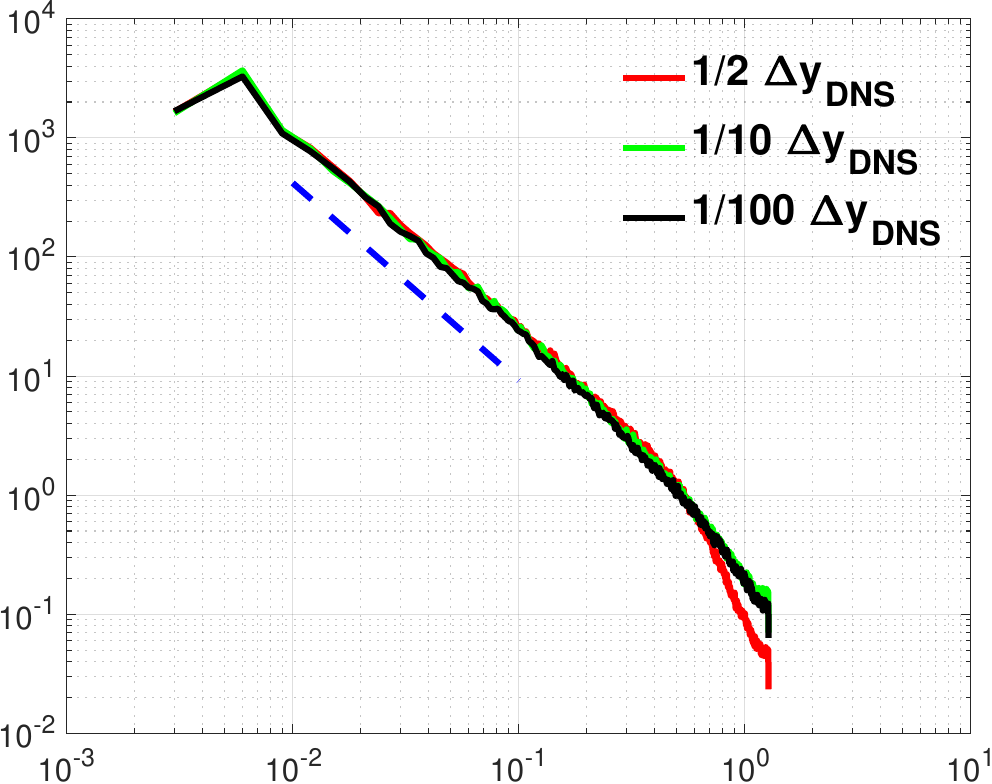}};
  \node[black] (B) at ($(A.south)!-.07!(A.north)$) {$k\eta$};
  \node[black,rotate=90] (C) at ($(A.west)!-.01!(A.east)$) {$E_{{22}}(k)/(\epsilon\nu^{5})^{1/4}$};
\end{tikzpicture}
\end{center}
\caption{Normalized ODT one-dimensional spectra of transverse velocity fluctuations for homogeneous decaying turbulence at $Re_{\lambda_{g}} = 500$. $E_{{22}}(k)/(\epsilon\nu^{5})^{1/4}$ (solid lines); $(k\eta)^{-5/3}$ (dashed blue line).} 
\label{spectrum500} 
\end{figure}

\begin{figure}[ht!]
        \centering
  \begin{tikzpicture}
  \node[inner sep=0pt] (A) {\includegraphics[width=0.4\textwidth]{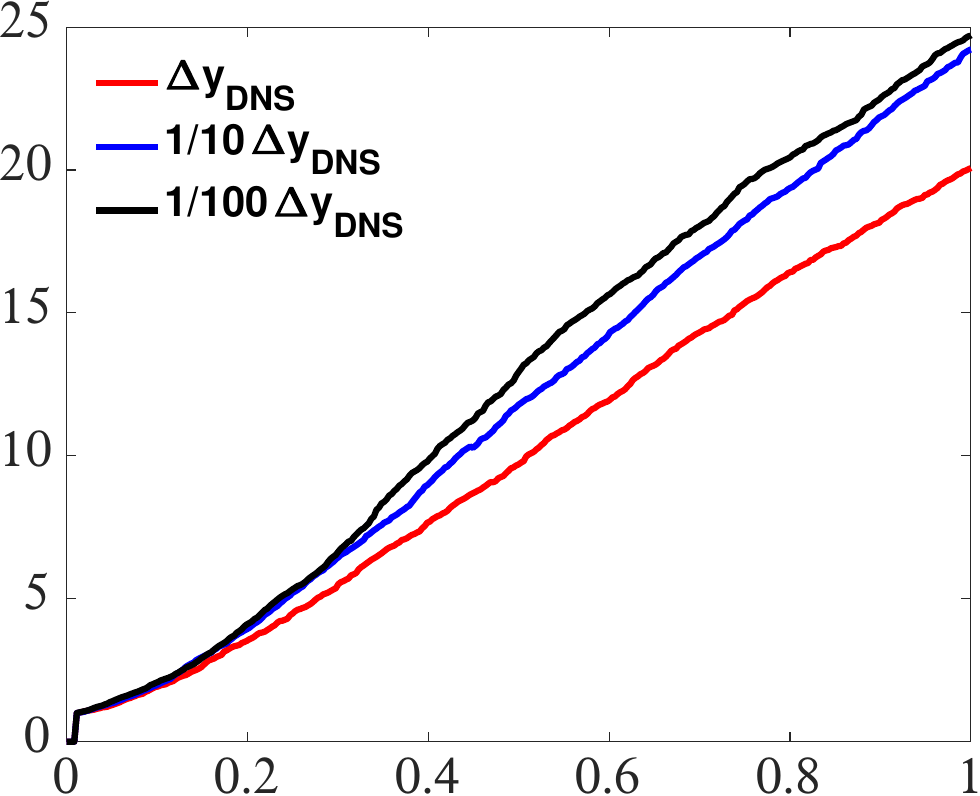}};
  \node[black] (B) at ($(A.south)!-.07!(A.north)$) {$t/\tau$};
  \node[black,rotate=90] (C) at ($(A.west)!-.07!(A.east)$) {Mean number of interfaces};
\end{tikzpicture}      
\caption{Mean number of interfaces predicted by ODT at $W\!e_{\lambda_{g}} = 100$ and $Re_{\lambda_{g}} = 500$.}
\centering
\label{MeshStudy500}
\end{figure}

Interface resolution is thus seen to drive the cost of the simulation for conditions relevant to subgrid closure, requiring a scale range up to five orders of magnitude. ODT simulation under these conditions is too costly to operate as a fully coupled closure model in each VOF mesh cell, but it is feasible to run such ODT cases off-line in order to build a look-up table that can be used for VOF closure.

\section{Parameter dependencies of the critical length scale} \label{sec:crit}

The separation scale of neighboring interfaces is seen to be both an important signature of turbulence interaction with surface tension and an important determinant of mesh resolution requirements. It is termed the critical radius in the context of droplet breakup \cite{gorokhovski2001breakup}, so for present purposes it is termed the critical scale. In Appendix \ref{sec:analysis}, prior analysis of the parameter dependencies of this scale is restated in a way that identifies a dimensionless group that is predicted to collapse those dependencies onto a universal scaling function based on the Kolmogorov phenomenology of homogeneous turbulence.

ODT critical length scales, DNS results, and the scalings derived in Appendix \ref{sec:analysis} are shown in Fig.~\ref{CriticalMap}. The critical length scale has been evaluated previously using DNS data \cite{mccaslin2014theoretical,mccaslin2015development} and the same method was used to obtain the DNS results in Fig.~\ref{CriticalMap}.

For each simulated ODT realization, the distances of the median interface location from its left and right nearest neighbors were deemed to be two estimates of the critical length scale. For a given case, these separation estimates were accumulated from all 2000 simulated realizations at time $t/\tau = 0.5$ and the median $l_{\sigma}$ of these estimates was taken to be representative of the ensemble value of the critical length scale. (The use of medians reduced the influence of outliers.) Thus, each critical-scale estimate is based on 4000 individual separation values.

\begin{figure}[ht!]
        \centering 
        \includegraphics[width=0.45\textwidth]{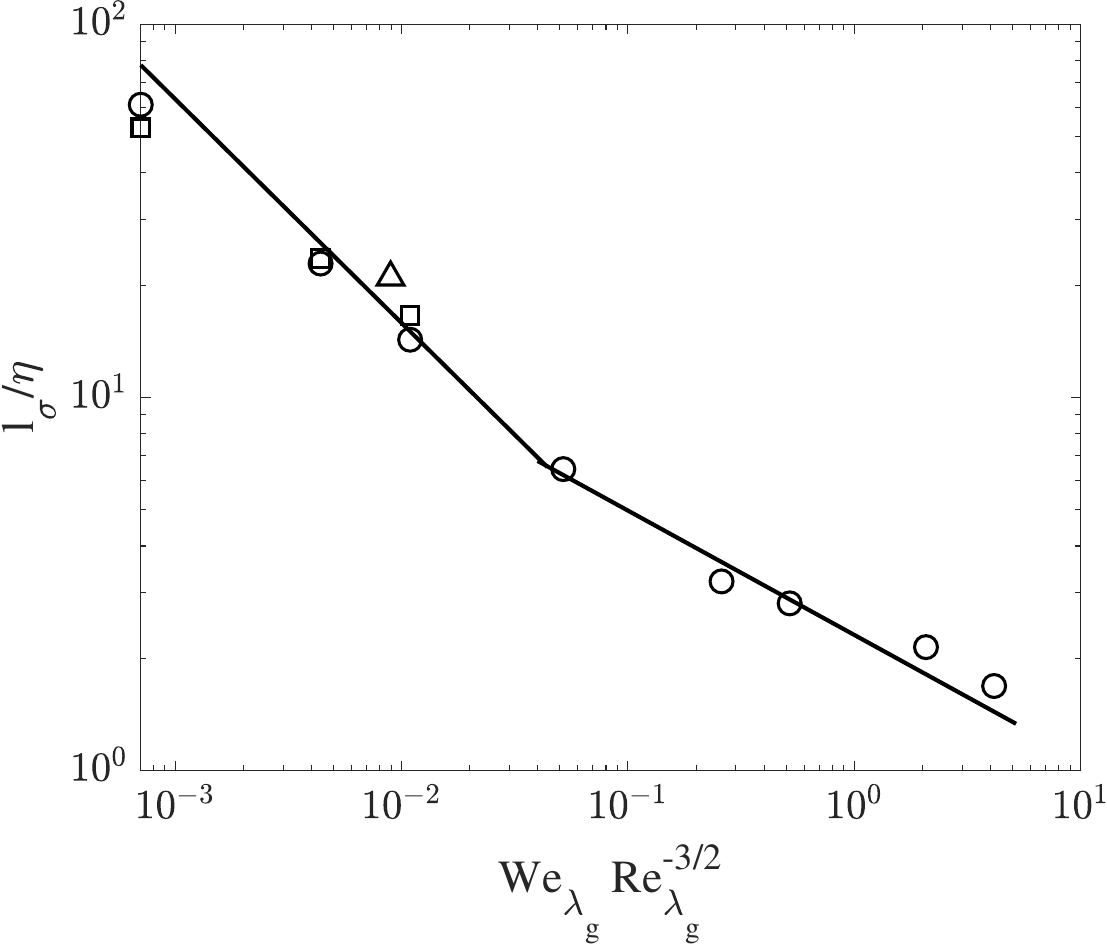}
\caption{Normalized ensemble-median critical length scale $l_{\sigma}/\eta$ plotted against the nondimensional group predicted to collapse all cases. $\bigcirc$, ODT cases 2-9; $\bigtriangleup$, ODT case 10; $\square$, DNS cases 2-4; $-\!\!\!-\!\!\!-$, predicted slopes $-\frac{3}{5}$ and $-\frac{1}{3}$.}
\centering
\label{CriticalMap}
\end{figure}

Several features of the results are noteworthy. First, ODT results are close to the DNS results notwithstanding the use of different methods to evaluate $l_{\sigma}$. Second, ODT results are consistent with theory for both scaling regimes. The theoretical scaling for the dissipative sub-range has not previously been confirmed and there apparently has been no confirmation for either sub-range that tested universality with respect to both Weber number and Reynolds number. Third, the application of the theory to the simulation results for this transient decaying case involves some inconsistency because the flow-state parameters used in the application of the theory are evaluated at $t=0$ rather than at $t = 0.5 \tau$, when $l_{\sigma}$ is evaluated. The agreement of simulation results with the theory despite this caveat suggests that the theory is more robust than implied by the assumptions on which it is best. However, the caveat advises caution in treating the results as numerical predictions, even with omitted coefficients restored.

To put this in context, ODT applications to turbulence modulation by dispersed particles indicated that consistent trends were obtained only if flow-state parameters were based on the particle-laden state rather than the initial state prior to the introduction of particles \cite{fistler_hit,fistler_hst}. (These flows remained homogeneous throughout the simulations, allowing consistent evaluation of flow-state parameters at any time.) Based on all noted consideration, there are several possible causes of the deviation of the high-Reynolds-number case (triangle in Fig.~\ref{CriticalMap}) from the overall trend, but any inconsistency with the Kolmogorov phenomenology that might be imputed would be mild at most. 

\begin{figure}[h!]
        \centering 
        \includegraphics[width=0.45\textwidth]{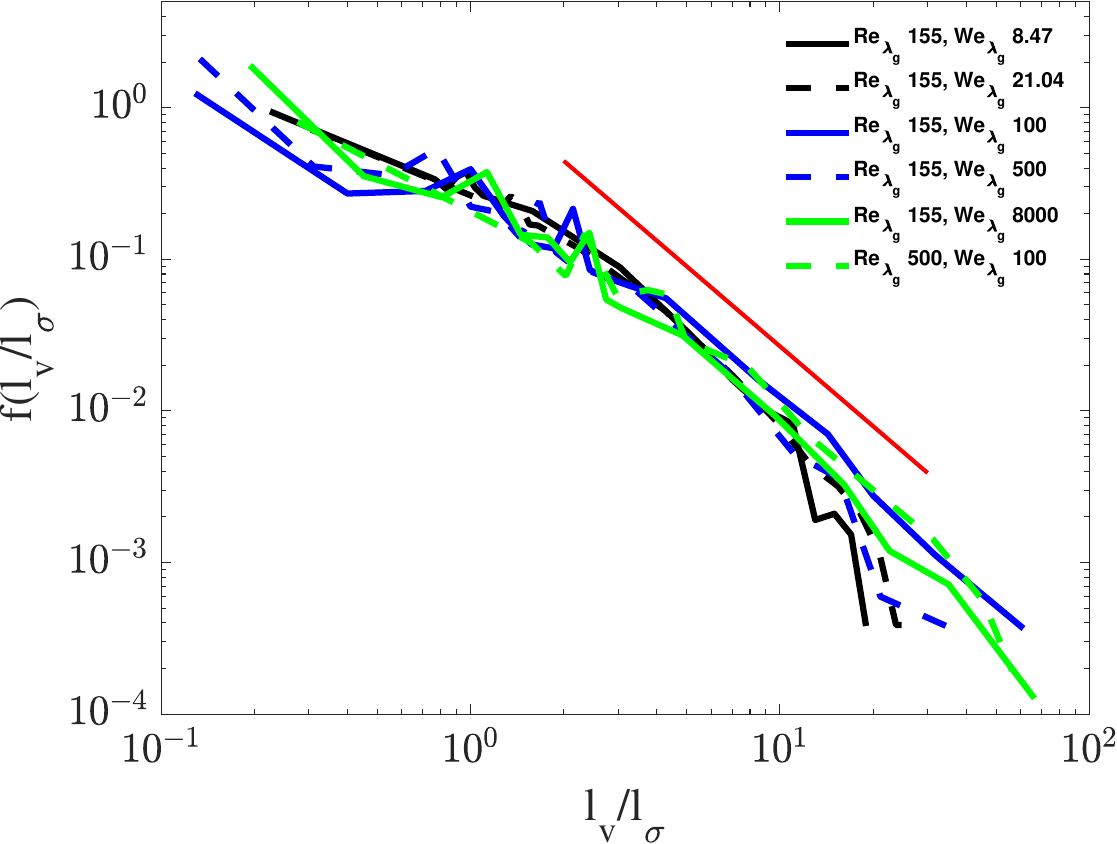};
\caption{For representative cases, the probability density function of interface separation values $l_v$ normalized by the ensemble-median value $l_{\sigma}$ for the case is shown. The red line of slope $-7/4$ demarcates a range of apparent power-law scaling.}
\centering
\label{CriticalDist}
\end{figure}

The 4000 individual separation values used for critical-distance estimation for a given case were additionally used to obtain the probability density function (PDF) of individual separation values $l_v$ normalized by the ensemble-median value $l_{\sigma}$ for the case. These PDFs are shown for selected cases in Fig.~\ref{CriticalDist}. These cases, spanning both of the turbulent-cascade regimes and including the high-Reynolds-number case, exhibit definitive collapse of the PDFs. Power-law falloff of the tails of the PDFs is seen, with an exponent value of roughly $-7/4$. Presently there is no evident explanation of the power-law falloff or the exponent value. Future investigation of the physical accuracy and phenomenological origins of these features is warranted.

The model is thus found to predict a significant degree of universality of the fluctuation statistics of the critical microscale. The simplicity of the model relative to Navier-Stokes turbulence suggests that this result is a bound on the degree of universality that might be obeyed physically. In any case, it raises the interesting question of how close the prediction is to physical reality. Addressing this will be challenging owing to the limitations of applicable computational and experimental techniques.

\section{Conclusions}

In previous work, ODT was extended for application to jet breakup. That extension included several empirical treatments designed to capture particular features of that configuration. Here, a simpler configuration requiring minimal empiricism and parameter adjustment has been studied in order to focus on interpretation and validation of the fundamental ODT representation of the interaction between surface tension and fluid inertia. To simplify the interpretation of results, the two fluid phases are assumed to have the same density and viscosity, so for vanishing surface tension, the phase interface becomes dynamically passive and the chosen flow configuration reduces to decaying homogeneous isotropic turbulence (HIT). Model parameters are set by comparing ODT and DNS results for this case and the model then predicts finite-$W\!e_{\lambda_{g}}$ behavior on this basis.

ODT and DNS results have been compared for a range of Weber numbers. For $W\!e_{\lambda_{g}} = \infty$, the interface is a passive material surface whose time development is itself useful validation information, demonstrating that ODT simulation of HIT captures key features of material surfaces advected by turbulence. As $W\!e_{\lambda_{g}}$ is reduced, the accuracy of the ODT predictions tends to decrease, reflecting the additional model simplifications associated with the treatment of surface-tension effects. Both by construction and by inference from comparisons to DNS results, the model is deemed to have useful predictive skill for the regime of high Weber and Reynolds numbers that is of greatest interest.

Accordingly, ODT application is extended beyond the DNS-accessible parameter space in order the evaluate the parameter dependencies of the Kolmogorov critical scale. Quantitative agreement with the available DNS results for this flow property lends credence to the broader set of results. Model results are consistent with the theoretical parameter dependencies in both the inertial and dissipative sub-ranges, where the latter does not appear to have been systematically investigated prior to the present study. Local fluctuation properties of the critical scale are then examined by evaluating PDFs of the separations of individual pairs of adjacent interfaces. Scaling the PDF for each considered case by the corresponding ensemble value, collapse of the PDFs onto a universal curve, encompassing both sub-ranges, is obtained. There does not appear to have been any prior investigation of such fluctuation properties. The present results suggest that this could be a fruitful avenue of future study.

\begin{acknowledgments}
The authors thank the Knut $\&$ Alice Wallenberg Foundation for financial support of this research.
\end{acknowledgments}

\appendix

\section{One-Dimensional Turbulence} \label{Sec:ODT}

\subsection{Time-advancement processes} \label{sec:advance}

ODT is a stochastic model simulating the evolution of turbulent flow along a notional line of sight through a three-dimensional flow. The flow state along that line of sight is treated as a closed system although physically it is an open system. This allows conservation laws to be enforced consistently, but not in the precise manner of 3D flow.

Time advancement of the present ODT formulation is expressed schematically as 
\begin{equation}
  {\dfrac{\partial u_{i}}{\partial t} - {\nu} \dfrac{\partial^{2}
  u_{i}}{\partial y^2}  = Eddies}, 
\label{momentum}
\end{equation}
where $\nu$ is the kinematic viscosity and the indices $i = 1$, 2, 3 denote the streamwise, lateral and spanwise velocity components, respectively, corresponding to the spatial coordinates $(x,y,z)$. This equation formally represents the two processes that can change the value of $u_{i}$ at a given location $y$ and time $t$. 

The left hand side represents viscous time advancement, which can be supplemented by body forcing and other case-specific processes. $Eddies$ denotes a model that idealizes the advective effect of 3D eddies, which are represented by instantaneous maps applied to property profiles, supplemented by energy redistribution among velocity components. Surface tension does not appear in Eq.~(\ref{momentum}) because its role in the model is to influence the occurrence and implementation of individual eddies.

Accordingly, turbulent advection is modeled in ODT by a stochastic sequence of events. These events represent the impact of turbulent eddies on property fields (velocity and any scalars that might be included) along the one-dimensional domain. During each eddy event, an instantaneous map termed the `triplet map,' representing the effect of a turbulent eddy on the flow, is applied to all property fields. It occurs within the spatial interval $[y_{0},y_{0}+l]$, where $y_{0}$ represents the eddy location on the ODT line and $l$ is the eddy size. A triplet map shrinks each property profile within $[y_{0},y_{0}+l]$, to one-third of its original length, inserts three identical compressed copies into the eddy range side by side so as to fill the range, and reverses the middle copy to ensure the continuity of each profile. The map mimics the eddy-induced folding effect and increase of property gradients. Formally, the new velocity profiles after a map are given by
\begin{align}
\hat{u}_{i}(y,t) = u_{i}(f(y),t),
\label{velocity treatment} 
\end{align}
here conveniently expressed in terms of the inverse map
\begin{align}
f(y) = y_{0} + 
\begin{cases}
3(y-y_{0}), & \text{if } y_{0}\leq y \leq y_{0}+{1 \over 3} l,
\\
2l - 3(y-y_{0}), & \text{if } y_{0}+{1 \over 3} l\leq y \leq y_{0}+{2 \over 3} l,
\\
3(y-y_{0}) - 2l, & \text{if } y_{0}+{2 \over 3} l\leq y \leq y_{0}+l,
\\
y-y_{0},         & \text{otherwise, }
\end{cases}
\end{align}  
which is single-valued. (The forward map is multi-valued.)

The triplet map is measure preserving, which implies that all applicable conservation laws 
are obeyed locally as well as globally. Various cases, such as buoyant stratified flow, 
involve sources and sinks of kinetic energy due to equal-and-opposite changes of one or more 
other forms of energy. Even in the simplest cases, viscosity converts kinetic energy into 
thermal energy. The resulting dissipation of kinetic energy is captured by 
Eq.~(\ref{momentum}). Energy-conversion mechanisms other than viscous dissipation 
are incorporated by introducing an additional operation during the eddy event. The 
implementation of this operation to account for surface-tension changes is described.

In the present formulation, the triplet map can increase the number of phase interfaces 
within the eddy interval, as illustrated in Section \ref{sec:multi}, resulting in an increase 
$\Delta E_{\sigma}$ 
of surface-tension energy that must be balanced by an equal-and-opposite decrease of 
kinetic energy $\Delta E_{kin}$, such that the total eddy-induced energy change 
$\Delta E = \Delta E_{kin} + \Delta E_{\sigma} $ is zero. 

Accordingly, the formal statement of the eddy-induced flow change in 
Eq.~(\ref{velocity treatment}) is generalized to
\begin{align}
  {{{\hat{u}_{i}}}}(y,t) = u_{i}(f(y),t) + c_{i}K(y) + b_{i}J(y).
\label{velocity treatment {{kernel}}} 
\end{align}
Here, $K(y) \equiv y - f(y)$ is the map-induced displacement of the fluid parcel that 
is mapped to location $y$ and $J(y) \equiv \arrowvert K(y) \arrowvert$.

The six coefficients $b_i$ and $c_i$ are evaluated by enforcing the prescribed 
kinetic-energy change based on the surface-tension energy change $E_{\sigma}$, 
which is zero if the eddy interval contains only one phase. Momentum conservation 
in each direction $i$ implies three more constraints. The two additional needed 
constraints are obtained by modeling the eddy-induced redistribution of kinetic energy 
among the velocity components. In accordance with return-to-isotropy phenomenology, 
these additional constraints are configured to impose a tendency of the component kinetic energies to equalize.

Eddy events displace fluid elements and thus constitute a Lagrangian representation 
of turbulent advection. The corresponding Eulerian interpretation in terms of Eq.~(\ref{momentum}) 
is that each event corresponds to an instantaneous change of properties at given $y$, 
so $Eddies$ is a sum of delta functions in time with weights that each represent the 
event-induced change of $u_i$ at location $y$ at the time of occurrence of a given event. 
This interpretation defines a formal representation of ODT time advancement in terms of 
Eq.~(\ref{momentum}), but the Lagrangian fluid-displacement picture is more intuitive 
and closer to the numerical implementation of the model.

\subsection{Eddy selection} \label{sec:selec}

ODT samples eddy events from an event-rate distribution that depends on the 
instantaneous flow state and therefore evolves with the flow. Thus, there is 
neither a predetermined frequency of occurrence of eddy events collectively nor of a particular 
eddy type corresponding to a given location $y_0$ and size $l$.
 
{{The mean number of events during a time increment $dt$ for eddies located within the interval 
$[y_{0},y_{0}+dy]$ in the size range $[l,l+dl]$ is denoted $ \lambda(y_{0},l; t) \, dy_{0} \, dl \, dt$. The relation
\begin{align}
\lambda(y_{0},l; t) = C/(l^{2}\tau(y_{0},l; t))
\label{velocity treatment {{lambda}}}
\end{align}
defines an eddy time scale $\tau$ and an adjustable parameter $C$ that scales the overall eddy frequency, 
{\color{black} where the argument $t$ appearing on both sides of the equation indicates that both $\lambda$ 
and $\tau$ vary with time for given values of $y_0$ and $l$ because $\tau$ depends on the time-varying 
instantaneous flow state in the manner described next. (With this understanding, the arguments of $\tau$ are 
henceforth suppressed.)} 
The dimensions of the event-rate distribution $\lambda$ are (length$^2 \times$ time)$^{-1}$.
To find {{the}} eddy time scale $\tau$, the square of the implied eddy velocity $l / \tau$ is modeled as{{}}
\begin{align}
  (l/\tau)^{2}  \sim E_{final} - Z(\nu^{2}/l^{2}) ,
\label{velocity treatment {{energy}}} 
\end{align}
where the first term, which is dependent on the instantaneous flow state, is specified by 
Eq.~(\ref{eq:eFinal}) in Section \ref{sec:multi} and the second term involving the parameter $Z$ suppresses 
unphysically small eddies. The coefficient implied by the proportionality is absorbed into $C$.

In practice it would be {{computationally unaffordable}} to reconstruct the event-rate 
distribution every time an eddy event or {{an advancement of Eq.~(\ref{momentum})}} takes place. 
{{Therefore eddy events are sampled using an}} equivalent Monte-Carlo numerical procedure called 
{{thinning}} \citep{ross1996stochastic}.

\subsection{Multiphase eddy implementation in ODT} \label{sec:multi}

If the eddy range contains one or more phase interfaces, then $\Delta E_{\sigma}$ must be evaluated in order 
to incorporate the surface-tension effect on eddy implementation as described in Section \ref{sec:advance}.
The procedure is motivated by Fig.~\ref{Multiphase}.a, which shows the initial state within an eddy that
contains a phase interface. For discussion purposes, the phases are termed liquid and gas although they 
have the same densities and viscosities in this study.
\begin{figure}[ht!]
	\centering
  \includegraphics[width=9cm]{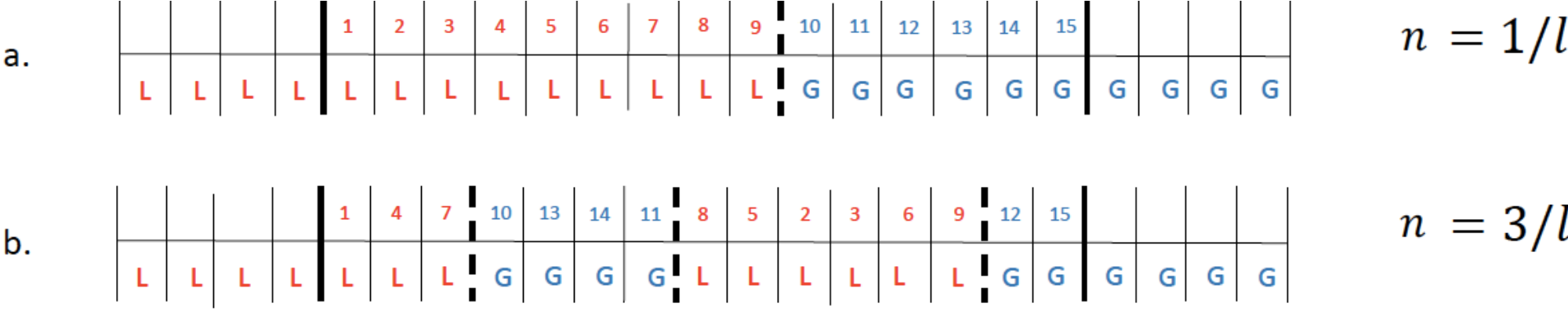}
\caption{Multiphase eddy treatment in ODT. (a) The size-$l$ spatial interval between 
the thick solid lines is selected for eddy implementation. It is a multiphase eddy
containing both liquid (L) and gas (G) separated by one phase interface (thick
dashed line), corresponding to interface number density $n = 1/l$ within the eddy interval. 
(b) A triplet map is implemented here as 
a permutation of the cells of a uniform spatial discretization of the 1D domain, 
illustrated by the reordering of cell indices within the eddy. Now there are three 
phase interfaces, corresponding to $n = 3/l$ and thus $\delta = 2/l$. }  
\label{Multiphase} 
\end{figure}
This eddy is energetically allowed only if there is sufficient kinetic energy available 
to supply the surface-tension energy needed to create the amount of new interface 
resulting from map implementation. Modeling is needed in order to specify the 
surface-tension energy change.

Namely, the ODT analog of the volumetric 
density $\sigma\alpha$ of surface-tension energy,
where $\sigma$ is the surface-tension energy per unit
area and $\alpha$ is the surface area per unit volume, 
must be identified, corresponding to the energy density
\begin{align}
E_{\sigma} = \sigma\alpha / \rho
\end{align}
per unit mass.  The meaning and
evaluation of $\alpha$ in ODT are considered.

Since an interface in ODT is represented by an isolated point on a line, geometric
interpretation is required in order to obtain the area increase implied by, e.g., the 
triplet map illustrated in Fig.~\ref{Multiphase}.  A plausible assumption for highly turbulent 
cases involving wrinkled interfaces is that the interface is a statistically homogeneous 
isotropic random surface. For such a surface, the number density $n$ of interface intersections 
along a line of sight corresponds to an interface area per unit volume $\alpha = 2n$ 
\citep{chiu2013stochastic}. This assumption is not precisely accurate for cases of interest, 
but it is convenient to adopt it as a universal assumption rather than to attempt 
a case-by-case treatment. On this basis,
\begin{equation}
E_{\sigma} = 2n\sigma / \rho,
\end{equation}
where in this context, $n$ denotes interface number density within the eddy interval, 
i.e.\ the number interfaces in the interval divided by $l$.

In Fig.~\ref{Multiphase}.a, $n = 1/l$ initially.
Triplet mapping of a phase interface within an eddy produces three such 
interfaces. This is shown in Fig.~\ref{Multiphase}.b and is interpreted as a
 tripling of interfacial area. The eddy-induced increase $\delta$ of the number 
 density of interfaces due to triplet mapping is thus $2/l$ for this eddy. Based on the relation 
 $\alpha = 2n$, the interfacial area increase per unit volume is $2 \delta$. Multiplication by the 
 surface tension $\sigma$ gives the surface tension potential energy per unit volume 
 that is stored in the newly created interfaces. This implies the surface tension 
 energy change per unit mass
 \begin{equation} \label{esigma}
\Delta E_{\sigma} = 2\sigma\delta/ \rho.
\end{equation}
This explanation corrects a previous \cite{movaghar2017numerical} erroneous discussion 
of these points, but the final result, Eq.~(\ref{esigma}), is unchanged.

Conservation of total energy requires an equal and opposite change of the final 
kinetic energy. Here this implies 
\begin{equation}
E_{final} = E_{kin} - \Delta E_{\sigma},
\label{eq:eFinal}
\end{equation}
where $E_{kin}$ and $E_{final}$ are the available kinetic energy per unit mass 
before and after the change, respectively. {\color{black} Here, available means the maximum 
amount extractable by adding weighted $J$ and $K$ kernels to the instantaneous velocity profiles 
as shown in Eq.~(\ref{velocity treatment {{kernel}}}). The change is implemented by similarly 
modifying the velocity profiles using weighted $J$ and $K$ kernels, but in this instance constraining 
the weighting coefficients $b_i$ and $c_i$ so as to extract the energy $\Delta E_{\sigma}$ from the 
flow field within the eddy interval.

If this procedure yields negative $E_{final}$, then the eddy is deemed to have insufficient available kinetic energy to overcome the surface-tension energy sink, so the eddy is not implemented. For positive $E_{final}$, the viscous term in Eq.~(\ref{velocity treatment {{energy}}}) might cause the right-hand side of that equation to be negative, so the viscous damping suppresses eddy implementation. Otherwise, the eddy is eligible to be selected for implementation based on Eq.~(\ref{velocity treatment {{lambda}}}).

\section{Critical-scale analysis} \label{sec:analysis}

An eddy of size $l$ with characteristic velocity fluctuation $u_l$ is prevented from overturning an interface if the inertia at scale $l$ is dominated by surface tension, hence if the Weber number $We = \rho u_l^2 l /\sigma$ is less than order unity. (Here and below, numerical coefficients are omitted.) The critical length scale $l_{\sigma} \sim \sigma / \rho u_l^2$ follows from this simple balance. The parameter dependence of $u_l$ will depend on where the critical length scale falls within the turbulent cascade sub-ranges. On this basis, the expressions for the critical length scale in the inertial and dissipation sub-ranges are found to be $l_{\sigma} = (\sigma^{3} /(\rho^{3} \epsilon^{2}))^{1/5}$ and $l_{\sigma} = (\sigma \nu /(\rho \epsilon))^{1/3}$, respectively \cite{kolturbulence,gorokhovski2001breakup}.

Here, the reasoning is restated in a way that allows both regimes to be expressed in terms of one dimensionless group. First suppose that $l_{\sigma} < \eta$, corresponding to the dissipative sub-range. At scales below $\eta$, the strain rate $\gamma = (\epsilon / \nu)^{1/2}$ at scale $\eta$ is applicable in this sub-range.

Surface tension resists the straining effect at a scale $l_{\sigma}$ at which the local Weber number $\gamma^2 l_{\sigma}^3 \rho / \sigma$ is of order unity, so nominally 1. This gives $l_{\sigma} = \left[ \sigma / (\gamma^2 \rho) \right]^{1/3}$. This estimate is valid provided that the surface tension does not suppress any inertial-range motions, hence for $l_{\sigma} < \eta$. The crossover to the inertial-range case corresponds to $l_{\sigma} = \eta$, giving $\left[ \sigma / (\gamma^2 \rho) \right]^{1/3} = Re^{-3/4} L$ based on the inertial-range scaling $\eta = Re^{-3/4}L$. Substitution of the inertial-range scaling $\gamma = (L / \eta)^{2/3}( u_{rms}/L) = Re^{1/2} (u_{rms}/L)$,  gives $\left[ L^2 \sigma / (u_{rms}^2 \rho Re) \right]^{1/3} = Re^{-3/4} L$.
In terms of the integral-scale Weber number $W\!e = u_{rms}^2 L \rho / \sigma$, this gives $(W\!e  Re)^{-1/3} = Re^{-3/4}$, which implies that $W\!e = Re^{5/4}$ is the crossover condition. Hence the governing dimensionless group at all scales is $W\!e Re^{-5/4}$.

To convert this to Taylor-scale properties, the starting point is the definition $\lambda_g = \eta^{2/3}L^{1/3} = (\eta /L)^{2/3}L = Re^{-1/2}L$, again omitting the numerical coefficient. Since $Re$ and $W\!e$ are both linear in $L$, this gives $Re_{\lambda_g} = Re^{1/2}$ and $W\!e_{\lambda_g} = Re^{-1/2} W\!e = Re_{\lambda_g}^{-1} W\!e$. Substitutions into $W\!e Re^{-5/4}$ then give $Re_{\lambda_g} W\!e_{\lambda_g} Re_{\lambda_g}^{-5/2} = W\!e_{\lambda_g} Re_{\lambda_g}^{-3/2}$, which is expresses the dimensionless group in terms of Taylor-scale properties.

The balance between kinetic energy and the surface-tension kinetic-energy barrier that determines $l_{\sigma}$ is conveniently expressed as $\rho u_{\sigma}^2 = \sigma / l_{\sigma}$, where $u_{\sigma}$ is the velocity fluctuation at scale $l_{\sigma}$. In terms of a local Weber number $W\!e_{\sigma} \equiv  u_{\sigma}^2 l_{\sigma} \rho / \sigma$, this corresponds to $W\!e_{\sigma} = 1$. For each regime of interest, the parameter dependence of $u_{\sigma}$ must be specified.

For the inertial sub-range, $u_{\sigma}$ is the eddy velocity at scale $l_{\sigma}$, hence $u_{\sigma} = (l_{\sigma} / L)^{1/3}u_{rms}$. Then the balance condition gives $\rho (l_{\sigma} / L)^{2/3}u_{rms}^2 = \sigma / l_{\sigma}$, which reduces to $l_{\sigma} / L = [\sigma / (u_{rms}^2 L \rho)]^{3/5} = W\!e^{-3/5}$. Finally, $l_{\sigma} / \eta = (L / \eta) W\!e^{-3/5} = Re^{3/4} W\!e^{-3/5} =(W\!e Re^{-5/4}) ^{-3/5} = (W\!e_{\lambda_g} Re_{\lambda_g}^{-3/2}) ^{-3/5}$. Normalization of $l_{\sigma}$ by $\eta$ sets the scaled crossover value to unity so that a deviation from this result conveniently quantifies the effect of neglecting numerical coefficients.

For the dissipative sub-range, the velocity fluctuation at scale $l_{\sigma}$ is $l_{\sigma}$ times the shear $\gamma$ at scale $\eta$, hence $u_{\sigma} = l_{\sigma} \gamma$. The balance condition then gives $l_{\sigma} / \eta = [\sigma / (\gamma^2 \rho)]^{1/3} / \eta = [\sigma / (\gamma^2 \eta^3 \rho)]^{1/3}$, which is the scale-$\eta$ Weber number to the $-1/3$ power. The scalings of $\gamma$ and $\eta$ in the inertial sub-range express $\gamma^2 \eta^3$ as $Re (u_{rms}/L)^2 Re^{-9/4} L^3 = Re^{-5/4} u_{rms}^2 L$.  The final result is $l_{\sigma} / \eta = (W\!e Re^{-5/4}) ^{-1/3} = (W\!e_{\lambda_g} Re_{\lambda_g}^{-3/2})^{-1/3}$.


\scalebox{0}{%
\begin{tikzpicture}
    \begin{axis}[hide axis]
        \addplot [
        color=red,
        solid,
        line width=2pt,
        forget plot
        ]
        (0,0);\label{hwplot1}
        \addplot [
        color=black,
        solid,
        line width=2pt,
        forget plot
        ]
        (0,0);\label{hwplot2}
         \addplot [
        color=green,
        solid,
        line width=2pt,
        forget plot
        ]
        (0,0);\label{hwplot3}
         \addplot [
        color=violet,
        solid,
        line width=2pt,
        forget plot
        ]
        (0,0);\label{hwplot4}
         \addplot [
        color=cyan,
        solid,
        line width=2pt,
        forget plot
        ]
        (0,0);\label{hwplot5}
        \addplot [
        color=blue,
        solid,
        line width=2pt,
        forget plot
        ]
        (0,0);\label{hwplotblue}
        \addplot [
        color=yellow,
        solid,
        line width=2pt,
        forget plot
        ]
        (0,0);\label{hwplotyellow}
         \addplot [
        color=red,
        dashed,
        line width=2pt,
        forget plot
        ]
        (0,0);\label{hwplot6}
    \end{axis}
\end{tikzpicture}%
}
\scalebox{0}{%
\begin{tikzpicture}
    \begin{axis}[hide axis]
        \addplot [
        color=red,
        mark=*,
        solid,
        line width=2pt,
        forget plot
        ]
        (0,0);\label{hwplot7}
        \addplot [
        color=black,
        mark= -,
        dashed,
        line width=1.5pt,
        forget plot
        ]
        (0,0);\label{hwplot8}
        \addplot [
        color=black,
        mark= --,
        dashed,
        line width=1.5pt,
        forget plot
        ]
        (0,0);\label{hwplot9}
         \addplot [
        color=green,
        mark= --,
        dashed,
        line width=2pt,
        forget plot
        ]
        (0,0);\label{hwplot10}
        \addplot [
        color=blue,
        mark= --,
        dashed,
        line width=2pt,
        forget plot
        ]
        (0,0);\label{hwplot11}
        \addplot [
        color=blue,
        solid,
        line width=2pt,
        forget plot
        ]
        (0,0);\label{hwplot12}
         \end{axis}
\end{tikzpicture}
}

\bibliography{ilass}

\end{document}